\documentclass[default, twocolumn]{aastex701}

\usepackage{soul, xcolor}
\usepackage{amsmath}

\newcommand{\rv}[1]{{\color{black} #1}}

\begin{document}

\title{Dynamical implications of the recently detected feature around Quaoar and constraints on the presence of additional satellites}

\author[0000-0001-5138-230X]{Gustavo~Madeira}
\affiliation{Observat\'orio Nacional/MCTI, R. General Jos\'e Cristino 77, CEP 20921-400, RJ, Brazil}
\email[show]{madeira@on.br}  

\author[0000-0001-5009-009X]{Leandro~Esteves}
\affiliation{UNESP - São Paulo State University, Grupo de Dinâmica Orbital e Planetologia, Av. Ariberto Pereira da Cunha, 333, Guaratinguetá, 12516-410, SP, Brazil}
\email[]{leandro.esteves@unesp.br}  

\author[0000-0003-0088-1808]{Bruno~E.~Morgado}
\affiliation{Universidade Federal do Rio de Janeiro - Observat\'orio do Valongo, Ladeira Pedro Ant\^onio 43, CEP 20080-090, RJ, Brazil}
\email{bmorgado@ov.ufrj.br}

\author[0009-0009-5523-5759]{Paulo~V.~S.~Soares}
\affiliation{UNESP - São Paulo State University, Grupo de Dinâmica Orbital e Planetologia, Av. Ariberto Pereira da Cunha, 333, Guaratinguetá, 12516-410, SP, Brazil}
\email[]{}

\author[0000-0002-3949-6045]{Silvia~M.~Giuliatti Winter}
\affiliation{UNESP - São Paulo State University, Grupo de Dinâmica Orbital e Planetologia, Av. Ariberto Pereira da Cunha, 333, Guaratinguetá, 12516-410, SP, Brazil}
\email[]{}  

\author[0000-0002-4901-3289]{Othon~C.~Winter}
\affiliation{UNESP - São Paulo State University, Grupo de Dinâmica Orbital e Planetologia, Av. Ariberto Pereira da Cunha, 333, Guaratinguetá, 12516-410, SP, Brazil}
\email[]{}  

\author[0000-0002-5960-6512]{Bruno~S.~Chagas}
\affiliation{Observat\'orio Nacional/MCTI, R. General Jos\'e Cristino 77, CEP 20921-400, RJ, Brazil}
\email[]{} 

\begin{abstract}
A recently reported opaque feature in the Quaoar system, detected during the 25 June 2025 stellar occultation, has been interpreted as either a $\sim 15$ km-radius satellite or a dense, sharp-edged arc orbiting at \rv{about $5600-5900$ km}. Here, we investigate both scenarios through numerical integrations that include Quaoar, its triaxial shape and Weywot. If the feature is a satellite, stability maps show that it acquires a forced eccentricity of about \rv{0.002} and has only a minor dynamical effect on the Q1R and Q2R rings. Its main effect is to clear a narrow region around its orbit, associated with the overlap of mean-motion resonances, \rv{preventing the long-term survival of kilometre-scale moons within $\sim$1 Quaoar radius of the satellite.} If instead the feature is an arc, we test confinement around the triangular equilibrium point of an unseen coorbital satellite. While azimuthal confinement is readily obtained for satellites with satellite-to-primary mass ratios up to $10^{-3}$, Weywot’s secular perturbation induces large radial excursions in the arc particles, preventing reproduction of the observed radial extent. The required radial confinement is achieved only for confiners comparable to or more massive than Weywot, which would likely be detectable by direct imaging. We therefore disfavour triangular-point confinement as an explanation for a long-lived dense arc and argue that the feature is more dynamically consistent with a satellite. \rv{We identify broad stable zones where additional undetected moons could reside, including near the rings as possible shepherd moons. Furthermore, Quaoar's ellipticity alone is sufficient to induce non-resonant orbital excitation capable of preventing ring coagulation, an effect that may be enhanced by small moons near the rings.}

\end{abstract}

\section{Introduction} \label{sec:intro}

The discovery of dense rings around Chariklo \citep{BragaRibas2014} changed our understanding of where planetary rings can exist, revealing that such structures are not restricted to giant planets. Subsequent discoveries of rings around Haumea, Quaoar, and Chiron \citep{Ortiz2017,Morgado2023,Pereira2023,Pereira2025} further indicate that the formation and long-term survival of rings may be relatively common among Centaurs and trans-Neptunian objects. Among these systems, Quaoar stands out as particularly puzzling, as it hosts \rv{two optically dense rings} located well beyond the classical fluid Roche limit, where orbiting material is expected to rapidly coagulate into satellites \citep{Morgado2023,Pereira2023}.

The persistence of these ring structures relies on the existence of radial confinement mechanisms \citep{GiuliattiWinter2023,Sickafoose2024} and on the occurrence of elastic impacts between ring particles. The latter is favoured by a combination of factors, such as the particle composition and porosity, low temperatures, and sufficiently high impact velocities \citep{Hatzes1988,Charnoz2011,Hyodo2015,Morgado2023}. In particular, the impact velocities arise as a direct consequence of the system’s dynamical environment, with the shape of the central body playing a key role in the evolution of orbiting material. Unlike the giant planets, which are nearly axisymmetric, minor bodies such as Quaoar exhibit significant azimuthal mass asymmetries that induce strong radial variations in nearby material \citep{Lages2017,Winter2019,Madeira2022a,Ribeiro2023,Madeira2025}.  

The non-axisymmetric shape of the central body -- in combination with its rotation -- gives rise to dynamically excited regions associated with spin–orbit resonances (SORs), which may act to confine ring material \citep{Sicardy2026,Salo2026} and possibly enhance impact velocities. This hypothesis, however, requires further and systematic investigation. Nevertheless, it does not appear to be a coincidence that the main rings of minor bodies are located close to low-order SORs. \rv{In the Quaoar system, the commensurability between the ring orbital periods and the spin period of the primary indicates that Q1R and Q2R may be under the influence of the 1:3 and 5:7 SORs, respectively \citep{Pereira2023}.}

In the specific case of Q1R, another mechanism that plays a role in the ring dynamics is the 1:6 mean-motion resonance (MMR) with Weywot, the only satellite currently confirmed \rv{to be orbiting} Quaoar. For realistic values of Weywot’s eccentricity \rv{(given its uncertainty, $0.056\pm0.093$)}, this resonance lies close to the ring, \rv{at a distance of about 29 km, and can increase the eccentricities of nearby particles} \citep{Rodriguez2023}.

\rv{This work also analysed, for the nominal value of Weywot’s eccentricity, the formation of clumps by assuming two values for the eccentricity of the ring particles (0.025 and 0.05). As expected, for the high value of eccentricity, dense clumps form in the ring. However, even for a lower value of eccentricity, a clump structure can also be observed. Therefore, the large azimuthal asymmetries observed in the optical depth of Q1R \citep{Morgado2023, Pereira2023} may be related to this process. However, further observations and a more realistic model, such as the shape of Quaoar and the solar radiation pressure, for example,  may confirm the influence of the 1:6 MMR with Weywot on the maintenance of the ring.}

A possibility that cannot be discarded is the action of additional satellites in exciting the ring particles and increasing their impact velocities. In this context, a pair of positive occultation chords reported in \cite{Nolthenius2025, Proudfoot2025a, BragaRibas2026} revealed the presence of an additional structure outside Q1R and interior to Weywot’s orbit. This new feature was discovered during a stellar occultation in June, 25$^{\rm th}$ 2025, when an opaque structure occulted the star for $\sim$1.2 seconds, translated into a chord of $\sim$30 km using the nominal velocity of the occultation ($24.5$ km/s). This event cannot be attributed to known member of Quaoar system (e.g. Quaoar itself, Weywot, Q1R and Q2R).

As detailed in \cite{Nolthenius2025} and \cite{BragaRibas2026}, this detection is intriguing: it was detected simultaneously in two different telescopes located in close range, and the star virtually disappeared, indicating a nearly opaque occulting body. Such a signature is consistent either with a previously unknown satellite around Quaoar or with a very dense arc of material. In the first scenario, the satellite is expected to have a physical radius of $15.1 \pm 0.6$ km and to orbit at a radial distance of $5676.5 \pm 0.9$ km \citep{BragaRibas2026}.

In the second scenario, the arc would orbit at the same radial location and have radial and azimuthal widths of $23 \pm 2$ km and $\lesssim 28^\circ$, respectively \citep{BragaRibas2026}. \rv{It is important to highlight, however, that the exact location of this feature is not well constrained. Because Quaoar's main body was not detected during the same occultation event, the inferred distance of the feature depends on the adopted ephemeris for Quaoar \citep[see also][]{Proudfoot2025a}.} 

Regardless of which scenario is correct, an important conclusion emerges: the Quaoar system is likely more complex than previously expected. \rv{The probability of detecting, via stellar occultation, a satellite with a radius of $\sim$15 km at this location is very low, leading \citet{BragaRibas2026} to speculate that a larger population of similar objects may exist.} Moreover, in the arc scenario, an additional confinement mechanism would be required, potentially involving one or several satellites to maintain such a structure \citep[see][]{Renner2014,Madeira2020,Madeira2022}.

In this work, we investigate the dynamical implications of the recently reported feature in the Quaoar system, seeking to shed light on whether it corresponds to a satellite or an arc of material. Given the realistic possibility that additional, yet non-detected, objects may exist in the system, we also map the stable regions where such bodies may reside and assess their implications for the rings. From here on, for clarity, we refer to the putative satellite associated with the reported feature -- or the coorbital satellite that might confine the putative arc -- as the “satellite”, and to any additional satellite that may reside in the system as a “moon”.

In Section~\ref{sec:constraint}, we derive observational and dynamical constraints on the Quaoar system, with emphasis on the ring eccentricities required to prevent coagulation. In Section~\ref{sec:methods}, we describe our numerical methods, while in Section~\ref{sec:moonlet} we perform numerical simulations of particle stability under the assumption that the feature is a satellite. In Section~\ref{sec:arc}, we examine the alternative scenario in which the feature corresponds to an arc confined by a single undiscovered satellite, characterising the properties that such a satellite would need to have in this scenario. Finally, in Section~\ref{sec:undiscovered}, we investigate the effects of additional undiscovered satellites in the system and present our conclusions in Section~\ref{sec:conclusion}.

\section{Dynamical and observations constraints on Quaoar system} \label{sec:constraint}

The Quaoar system imposes strong constraints on the possible presence of additional satellites. A defining characteristic of the system is that the massive Q1R ring lies well beyond the classical Roche limit for realistic values of bulk density, challenging standard views on the long-term stability of planetary rings. \citet{Morgado2023} showed that a sufficiently large radial velocity dispersion in the ring may prevent particle coagulation, a condition that could be sustained by gravitational excitation from additional satellites. In this context, we first focus on deriving constraints on the eccentricities of the Q1R ring particles.  

A first estimate of the particle eccentricities, based on the observational data of Q1R, can be obtained by adopting the simplest streamline model for the ring -- that is, by describing the particle trajectories as elliptical -- and assuming apsidal alignment of the particles. For such a configuration, the minimum ${\rm W_-}$ and maximum ${\rm W_+}$ radial widths of the ring are given by \citep{Nicholson1978,French1986,GiuliattiWinter2023}:
\begin{equation}
W_{\pm}=\delta a \pm a_r \delta e,
\end{equation}
where $a_r$ is the mean ring radius, $\delta a$ is the mean ring width, and $\delta e$ is the eccentricity gradient. Adopting \rv{$a_r = 4096$~km \citep{Proudfoot2025ring}}, ${\rm W_-} = 5$~km and ${\rm W_+} = 60$~km \citep{Morgado2023,Pereira2023}, and noting that the particle eccentricities are expected to be a few times $\delta e$, we infer eccentricities of $\gtrsim 10^{-2}$.

A second, independent estimate can be obtained by exploring the conditions under which the Q1R ring does not coagulate into a satellite. The classical Toomre parameter measures the gravitational stability of a ring and can be used as a tool to estimate the possible combinations of typical bulk density $\rho_p$ and typical physical radius $r_p$ of the particles, which are otherwise unknown parameters. The Toomre parameter $Q$ is given by \citep{Toomre1964,Goldreich1982}:
\begin{equation}
Q=\frac{3\Omega c_s}{4\pi G r_p \rho_p \tau}, \label{eqQ}
\end{equation}
where $G$ is the gravitational constant\rv{, $c_s$ is the random velocity dispersion of the ring particles,} $\Omega$ is the Keplerian frequency, and $\tau$ is the optical depth, here assumed to be $\tau = 0.4$, the maximum value measured for Q1R \citep{Morgado2023,Pereira2023}. For $Q>2$, a ring is gravitationally stable, while for $Q\lesssim 2$ wakes start to form within the ring, and for $Q\lesssim 1$ the ring is expected to coagulate \citep[see][]{Salmon2010}. 

\rv{In general, the random velocity dispersion $c_s$ depends on the thermal, physical, and orbital properties of the ring particles \citep[see][]{Goldreich1978,Shu1985}. Nevertheless, a crude estimate can be obtained by taking $c_s$ as the radial velocity associated with the particle eccentricity, $c_s \sim \bar{e}\, v_k$, where $\bar{e}$ is the mean eccentricity and $v_k$ the Keplerian velocity. Under this assumption, we can derive an order-of-magnitude estimate for the particle eccentricity required to avoid coagulation (i.e., $Q \gtrsim 1$).}

\begin{figure}
  \gridline{
    \fig{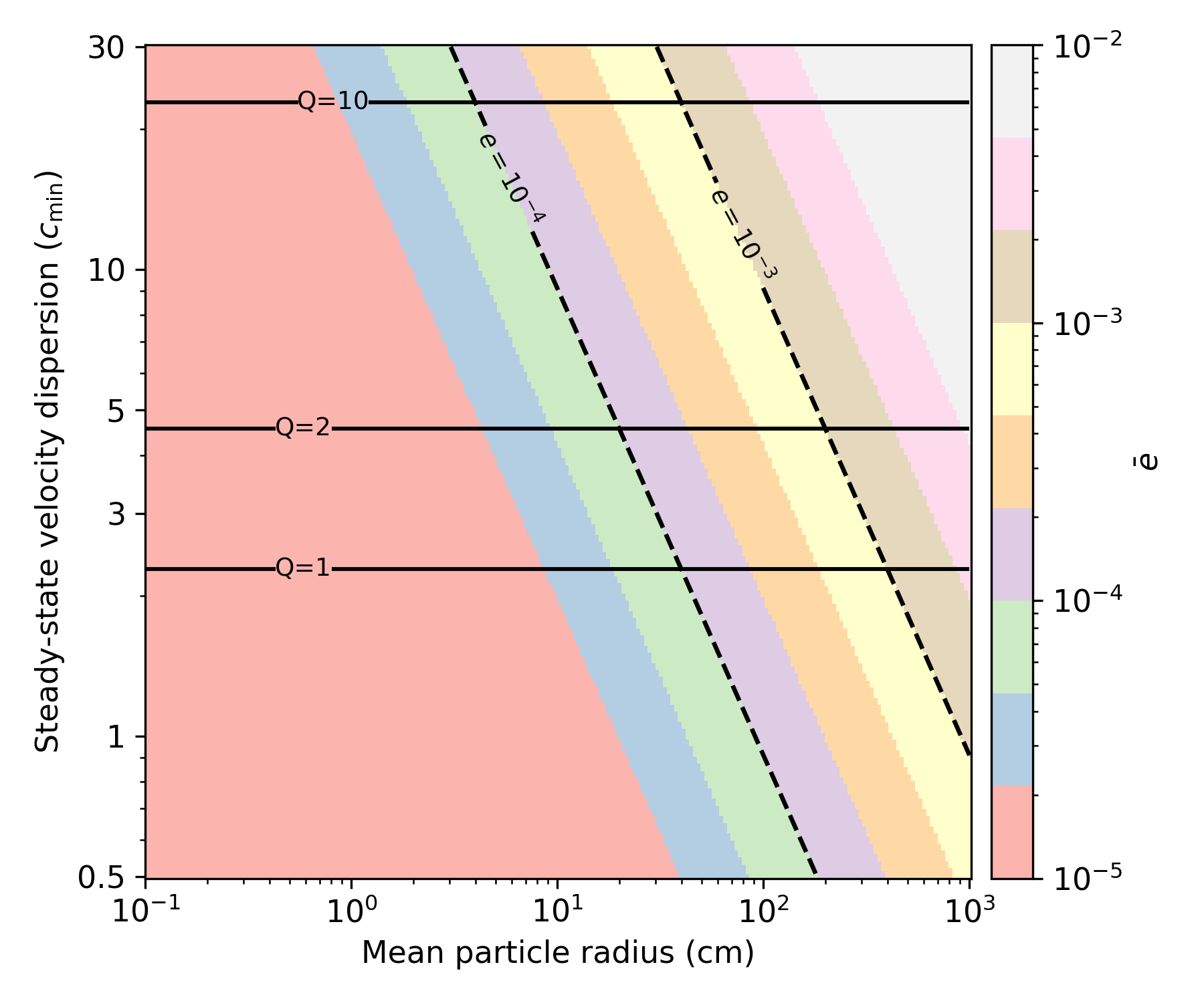}{\columnwidth}{(a) ${\rm \rho_p= 1000~kg\cdot m^{-3}}$}
    }
    \gridline{
    \fig{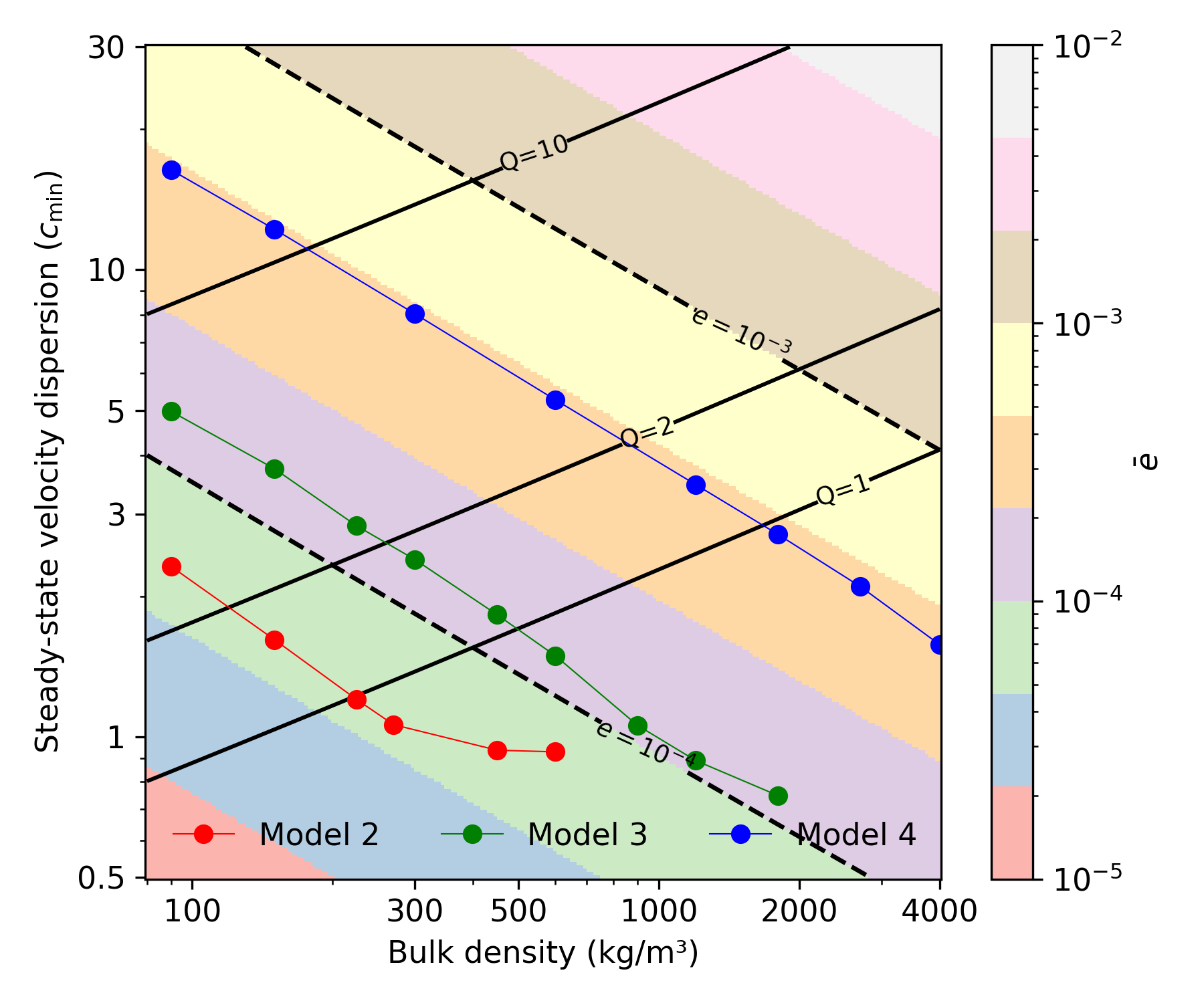}{0.5\textwidth}{(b) ${\rm r_p= 1~m}$}
  }
  \caption{\rv{Mean} particle eccentricity as a function of the steady-state velocity dispersion, physical radius, and bulk density of Q1R ring particles. In panel (a), the particle radius is varied over realistic values for a fixed bulk density of ${\rm 1000~kg\cdot m^{-3}}$, while in panel (b) the bulk density is varied and the particle radius is fixed at ${\rm 1~m}$. \rv{The colour scale represents the mean eccentricity, with the dotted lines indicating constant values of $\bar{e}$ and the solid black lines indicating constant values of the Toomre stability parameter $Q$.} Coloured symbols in panel (b) correspond to different collisional steady-state models from \citet{Morgado2023}.}
  \label{fig:eccentricities}
\end{figure}

\rv{Figure~\ref{fig:eccentricities} shows the velocity dispersion $c_s$ and the mean eccentricity of ring particles for different combinations of particle radius and bulk density. As a reference, we express the velocity dispersion $c_s$ in units of the minimum steady-state velocity dispersion $c_{\rm min}$ required to prevent accretion, as estimated by \citet{Morgado2023} through a set of box simulations with self-gravitating particles. The quantity $c_{\rm min}$ is given by \citep{Morgado2023}}
\begin{equation}
c_{\rm min}=1.65v_{\rm esc}+0.16v_{\rm esc}\left(\frac{\rho_p}{80~{\rm kg\cdot m^{-3}}}\right).
\end{equation}

\rv{Analysing the results of Figure~\ref{fig:eccentricities}a, we find that, regardless of particle size, a minimum velocity dispersion of $c_s \sim 5c_{\rm min}$ is required for the ring to remain gravitationally stable according to the Toomre criterion ($Q \gtrsim 2$). For $2.5c_{\rm min} < c_s < 5c_{\rm min}$, the ring remains in a particulate state but develops wake-like structures. Adopting the threshold value $Q = 1$, we find that a ring composed of icy particles with typical radii up to 40~cm must have eccentricities as low as $10^{-4}$ to avoid coagulation. For particles with radii of hundreds of metres, in turn, coagulation cannot be avoided, as eccentricities larger than unity would be required.}

\rv{Given the estimate by \citet{Morgado2023} that particles in Q1R with radii up to 0.2~cm undergo radial excursions exceeding the ring width due to solar radiation pressure, we infer that typical Q1R particle sizes range from tens of centimetres to a few metres -- this range is consistent with values inferred for the rings of Chariklo \citep{GiuliattiWinter2023}. For this size range, mean eccentricities of $\gtrsim 10^{-3}$ are sufficient to avoid coagulation.}

The $Q$ parameter scales with particle bulk density \rv{(Equation~\ref{eqQ})}, and lower eccentricities are sufficient to avoid coagulation for less dense particles \rv{(Figure~\ref{fig:eccentricities}b)}. For porous ice particles (${\rm \rho_p \sim 400–800~kg\cdot m^{-3}}$), \rv{mean eccentricities of $\sim 10^{-3}$ are also sufficient to ensure gravitational stability of the ring, corresponding to $c_s \gtrsim 3c_{\rm min}$.}

\rv{For comparison, we also show in Figure~\ref{fig:eccentricities}b the results from the numerical simulations of ring coagulation presented by \citet{Morgado2023}. These simulations explore different models for the coefficient of restitution in impacts, obtained empirically from laboratory experiments of collisions between icy particles \citep{Bridges1984,Hatzes1988}. Models~2 and 3 correspond to frost-covered ice particles of a few centimetres in size, at temperatures of 123~K and 210~K, respectively, while Model~4 represents $\sim$20~cm-sized particles with compacted-frost surfaces at 123~K.} 

\rv{As can be seen, the expected range of particle bulk density (${\rm \rho_p \sim 400–800~kg\cdot m^{-3}}$), physical radius (${\rm cm}$--${\rm m}$), and radial velocity dispersion ($c_s \gtrsim 3c_{\rm min}$) best matches the results of Model~4, which represents a ring composed of particles with these physical properties. This agreement provides support for the idea that Q1R is composed of particles with such properties. Given the exposed, we therefore adopt a benchmark eccentricity of $10^{-3}$} for the Q1R and Q2R ring particles when analysing the effects of the newly reported structure, as well as the possible presence of undiscovered satellites around Quaoar.

From an observational standpoint, only limited constraints can be placed on potential undiscovered structures or satellites in the Quaoar system. This is primarily because direct detections of satellites or rings in the trans-Neptunian region are not trivial. The Hubble Space Telescope (HST) has detected a significant number of binary systems; however, due to technical limitations, it is more efficient in observing large companions on wide orbits \citep{Noll2020}. 

In the specific case of the Quaoar system, even Weywot ($\sim$85~km in radius) can only be observed when it is separated from Quaoar by more than 0.2~arcseconds \citep[nearly 6000~km from Quaoar’s centre;][]{Fraser2010}. As pointed out by \cite{Proudfoot2025a}, even with the James Webb Space Telescope (JWST), a satellite with a radius of 15~km at a distance of $\sim$5600~km from Quaoar cannot be unambiguously identified. In this sense, it is possible that many moonlets with radii of a few tens of kilometres could coexist within the Quaoar system without having been previously detected.

\section{Numerical Methods} \label{sec:methods}
We investigate the dynamics of material orbiting Quaoar through numerical simulations performed with the \texttt{Rebound} package \citep{Rein2012}, using the IAS15 integrator \citep{Rein2015}. \rv{Quaoar is assumed to have a mass of ${\rm M_Q = 1.212 \times 10^{21}~kg}$ \citep{Proudfoot2025_5} and a spin period of ${\rm T_Q = 17.752~h}$ \citep{Kiss2024}. We model the object as a triaxial ellipsoid with mean physical radius ${\rm R_Q = 548.8~km}$ and second-degree gravitational coefficients ${\rm C_{20} = -0.043}$ and ${\rm C_{22} = 0.005}$, derived from the shape model presented in \cite{margoti2024,Proudfoot2025_5}.} Following the methodology of \citet{Madeira2025}, the equations of motion of an object around Quaoar ($\ddot{x}$, $\ddot{y}$, $\ddot{z}$) under the influence of its ellipsoidal shape, at simulation time $t$, are expressed in the quasi-inertial \texttt{Rebound} reference frame $(x, y, z)$ as follows:
\begin{equation}
\begin{aligned}
\ddot{x} &={} -\frac{GM_Q}{r^3}x 
+ \frac{GM_Q R_Q^2}{r^5}
\Bigg[
C_{20}\left(\frac{3}{2}x - \frac{15}{2}\frac{x z^2}{r^2}\right) \\
& + C_{22}\Bigg(
6x\cos\left(\frac{4\pi t}{T_Q}\right)
+ 6y\sin\left(\frac{4\pi t}{T_Q}\right) \\
& + \frac{15x}{r^2}
\left[
\left(y^2 - x^2\right)\cos\left(\frac{4\pi t}{T_Q}\right)
- 2xy\sin\left(\frac{4\pi t}{T_Q}\right)
\right]
\Bigg)
\Bigg],
\end{aligned}
\end{equation}

\begin{equation}
\begin{aligned}
\ddot{y} &={} -\frac{GM_Q}{r^3}y 
+ \frac{GM_Q R_Q^2}{r^5}
\Bigg[
C_{20}\left(\frac{3}{2}y - \frac{15}{2}\frac{y z^2}{r^2}\right) \\
& + C_{22}\Bigg(
6x\sin\left(\frac{4\pi t}{T_Q}\right)
- 6y\cos\left(\frac{4\pi t}{T_Q}\right) \\
& + \frac{15y}{r^2}
\left[
\left(y^2 - x^2\right)\cos\left(\frac{4\pi t}{T_Q}\right)
- 2xy\sin\left(\frac{4\pi t}{T_Q}\right)
\right]
\Bigg)
\Bigg],
\end{aligned}
\end{equation}

\begin{equation}
\begin{aligned}
\ddot{z} & ={} -\frac{GM_Q}{r^3}z 
+ \frac{GM_Q R_Q^2}{r^5}
\Bigg[
C_{20}\left(\frac{9}{2}z - \frac{15}{2}\frac{z^3}{r^2}\right) \\
& + C_{22}\frac{15z}{r^2}
\left[
\left(y^2 - x^2\right)\cos\left(\frac{4\pi t}{T_Q}\right)
- 2xy\sin\left(\frac{4\pi t}{T_Q}\right)
\right]
\Bigg].
\end{aligned}
\end{equation}
where $r=\sqrt{x^2+y^2+z^2}$ is the distance from Quaoar’s centre. Our simulations also include the satellite Weywot, with mass ${\rm M_w = 2.4 \times 10^{18}~kg}$ and physical radius ${\rm R_w = 85~km}$ \citep{Proudfoot2025_5}. \rv{The satellite is assumed to have a semi-major axis of ${\rm a = 13329~km}$ and eccentricity ${\rm e = 0.011}$ \citep{Proudfoot2025_5}, with all angular orbital elements set to zero.}

\rv{Despite evidence that the Weywot may have an inclination of $\sim5^{\circ}$ relative to the rings \citep{Proudfoot2025_5}, we assume it to orbit in Quaoar's equatorial plane, since the pole orientations of Quaoar and the rings are still poorly constrained. Nevertheless, tests including Weywot on a $\sim5^{\circ}$ inclined orbit show only a minor effect on the putative satellite and ring particles, inducing inclinations of $\ll1^{\circ}$ in both cases. This is expected given Weywot's large distance from these objects.}

Our stability simulations span ${\rm 10^4~T_Q}$, covering a period of about 20~years. Direct integration over timescales longer than a few decades is computationally costly; therefore, we assess the long-term stability of selected cases by computing the Lyapunov time $t_L$. In these selected cases, we rerun the simulations including only the surviving particles and assign to each of them a shadow particle with a small initial displacement in the $x$ coordinate of ${\rm \delta_0 = 10^{-6}~R_Q}$. Following \citet{Benettin1980,Wolf1985}, we compute, at regular intervals of ${\rm T_Q}$, the separation $\delta_k$ between the reference and shadow trajectories in phase space. The Lyapunov time is then estimated as
\begin{equation}
t_L^{-1} = \frac{1}{10^4 T_Q} \sum_{k=1}^{10^4} \ln\left(\frac{\delta_k}{\delta_0}\right).
\end{equation}

\section{The new feature as a satellite and its dynamical implications} \label{sec:moonlet}

In this section, we explore the feature reported by \cite{Nolthenius2025} and \cite{BragaRibas2026} as a satellite, following the interpretation of \cite{Proudfoot2025a}. This represents the simplest dynamical interpretation, as it does not require any additional confinement mechanism. To investigate the general stability of the system, we performed numerical simulations including an ellipsoidal Quaoar, Weywot, the putative satellite, and a set of massless test particles.

\rv{For our nominal simulation, 40000 particles are distributed over a grid in semi-major axis $a$ and eccentricity $e$, using 400 logarithmically spaced values of $a$ ranging from 1 to ${\rm 60~R_Q}$ and 100 values of $e$ ranging from ${\rm 10^{-4}}$ to 0.5. The putative satellite is assumed to have a physical radius of 15.1~km and bulk density ${\rm \rho = 1770~kg\cdot m^{-3}}$ \citep{BragaRibas2026}, corresponding to a mass of ${\rm \sim 2 \times 10^{-5}\,M_Q}$, about two orders of magnitude smaller than that of Weywot. It} is initially in a circular, equatorial orbit with a semi-major
axis of 5676~km \citep{BragaRibas2026}. The angular orbital elements of both the satellite and the test particles are set to zero.

\rv{To cover the scenario proposed by \citet{Proudfoot2025a}, we also performed a simulation with 10000 particles (a $100 \times 100$ grid) and with the putative satellite initially placed at a semi-major axis of 5910~km. Nevertheless, the results are qualitatively similar to the nominal case, and therefore only the nominal case is discussed below.} We emphasise that, when referring to orbital elements, we mean the osculating elements defined within the framework of the classical two-body problem. For a detailed discussion of orbital elements in the context of non-spherical bodies, we refer the reader to \cite{Ribeiro2021}.

\begin{figure*}
  \gridline{\fig{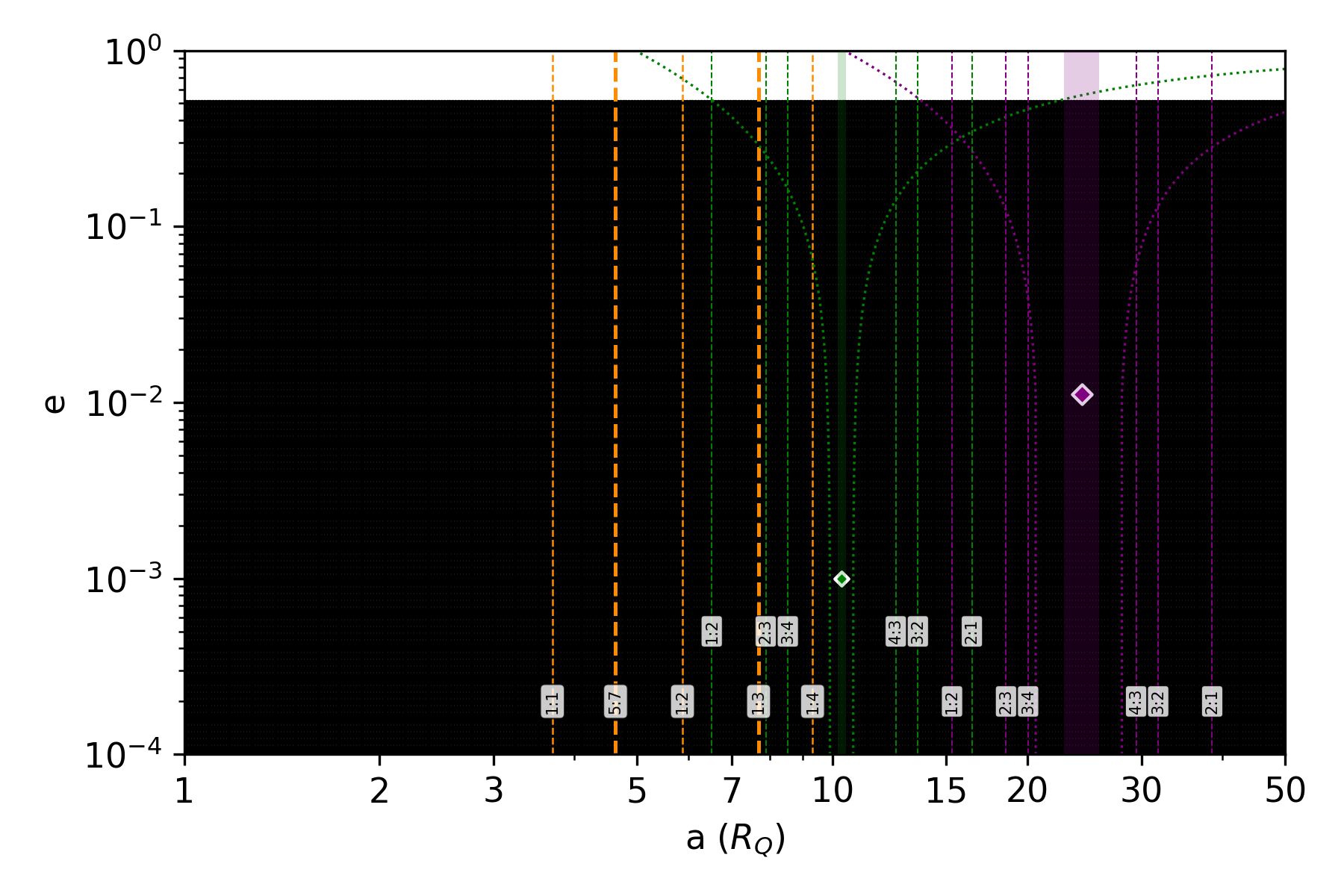}{0.49\textwidth}{(a) ${\rm 0}$}
    \fig{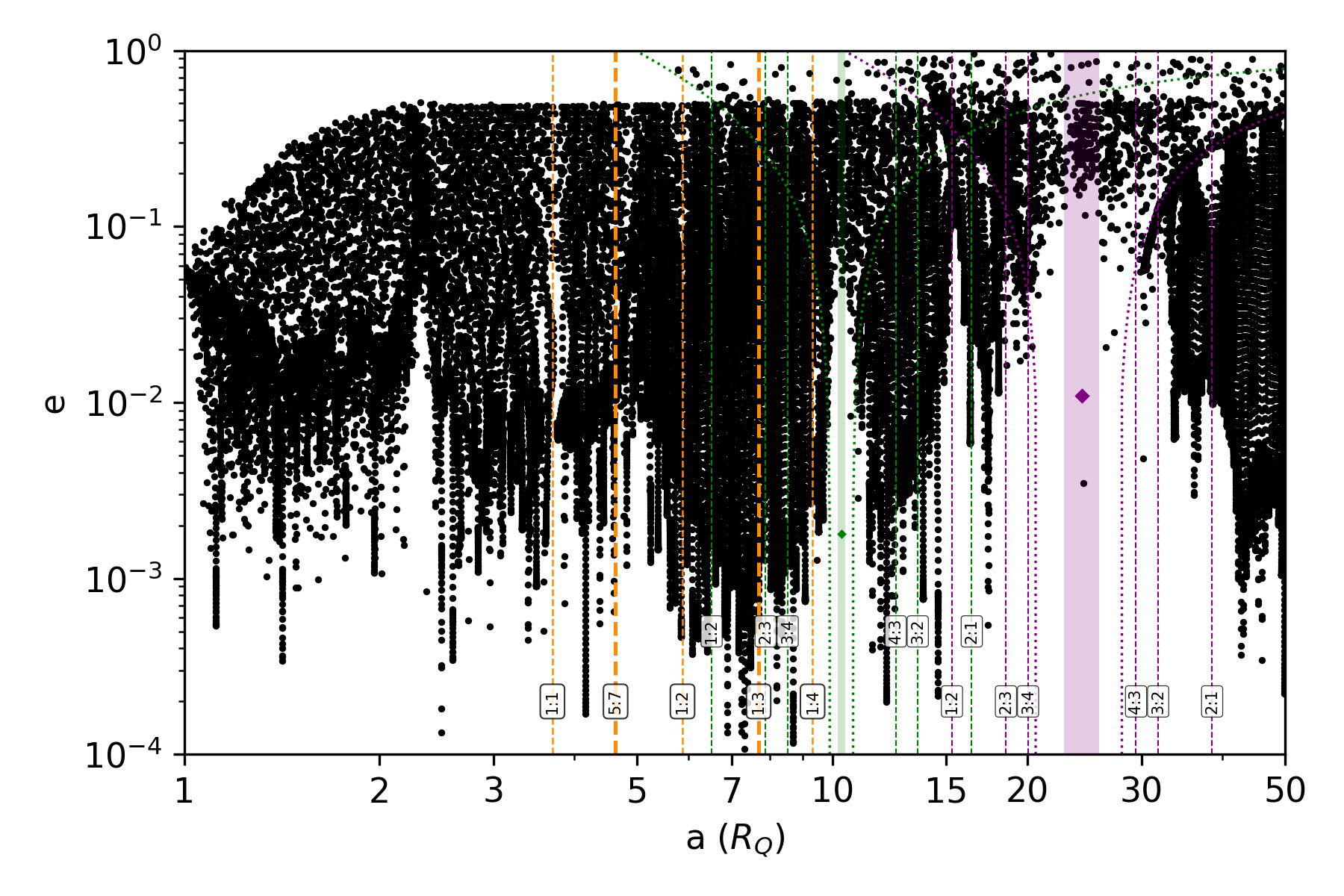}{0.49\textwidth}{(b) ${\rm 10^2~T_Q}$}
  }
  \gridline{
    \fig{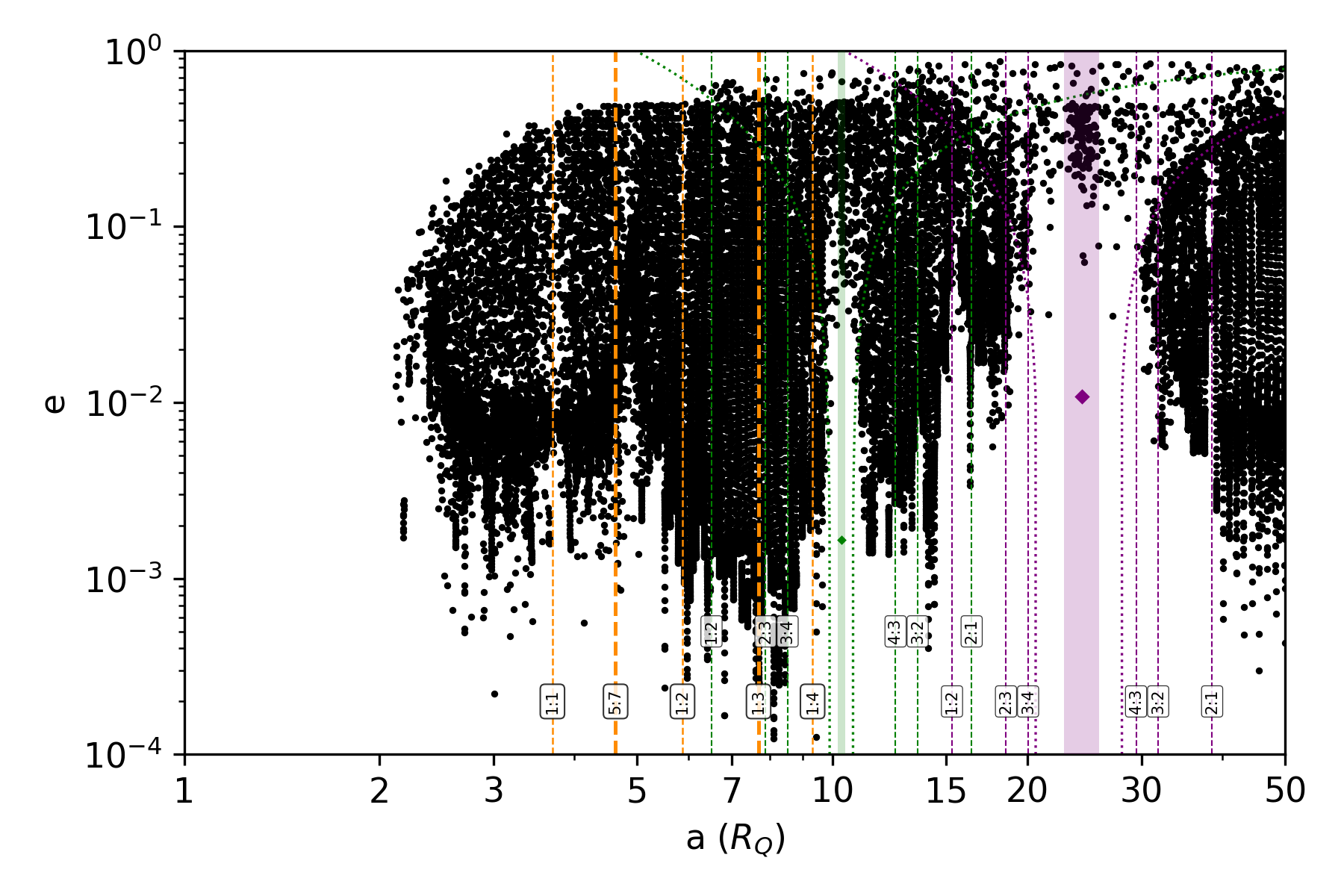}{0.49\textwidth}{(c) ${\rm 10^3~T_Q}$}
    \fig{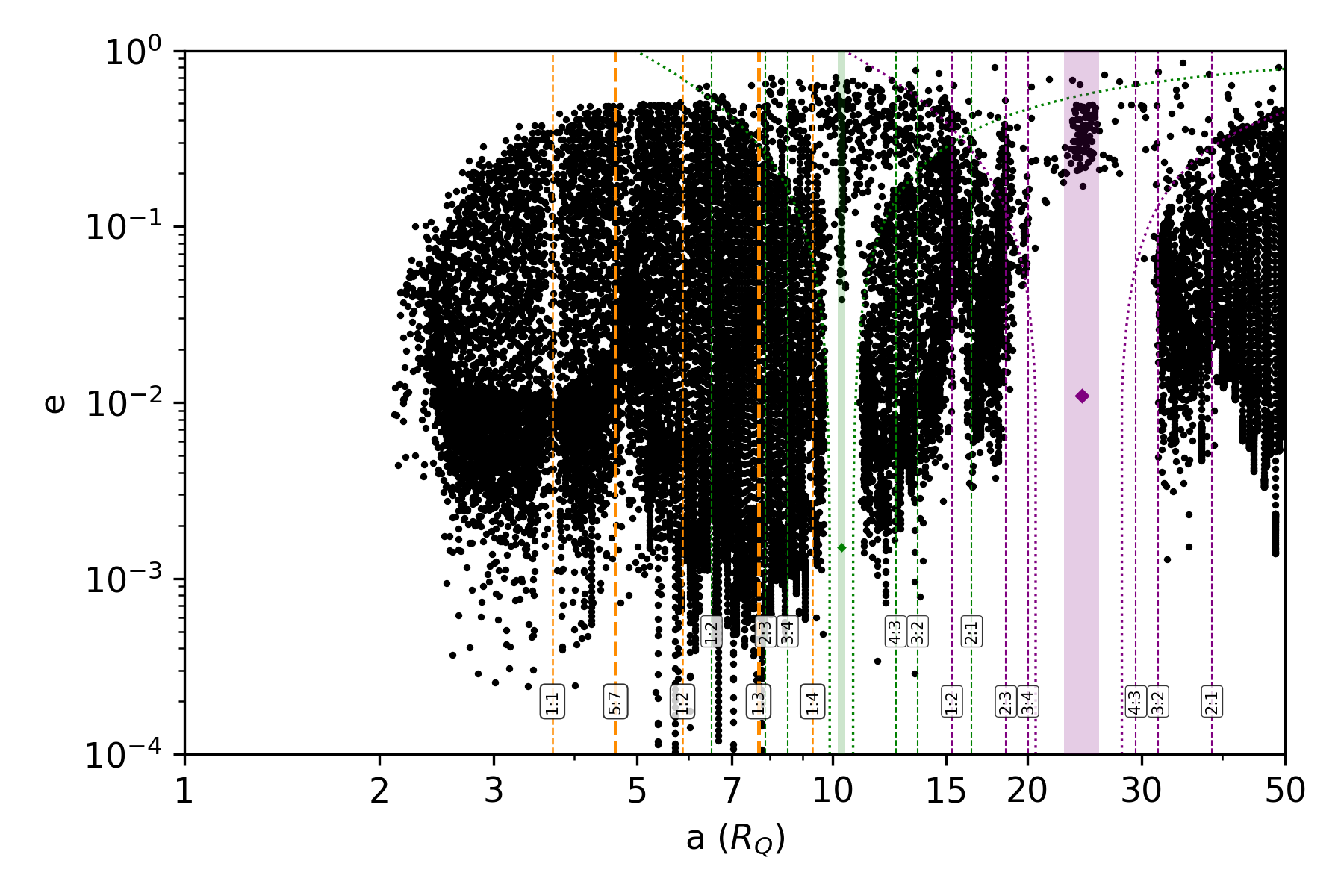}{0.49\textwidth}{(d) ${\rm 10^4~T_Q}$}
  }
  \caption{Evolution of eccentricity and semi-major axis of satellites (diamond markers) and test particles (black dots) around Quaoar, with each panel corresponding to a different epoch. Orange vertical dashed lines mark the locations of spin–orbit resonances with Quaoar, while green and purple dashed lines indicate first-order mean-motion resonances with the satellite (green diamond) and Weywot (purple diamond), respectively. Shaded vertical bands highlight the coorbital regions of each satellite, and the coloured curves delimit the region where particle apocentres and pericentres cross the chaotic zone surrounding the satellite, with the colours of the bands and curves matching those of the corresponding satellite. The wider dashed orange lines correspond to the SORs associated with the rings: the 1:3 SOR is associated with Q1R and the 5:7 SOR is associated with Q2R.}
  \label{fig:4panels}
\end{figure*}

\begin{figure*}
  \gridline{\fig{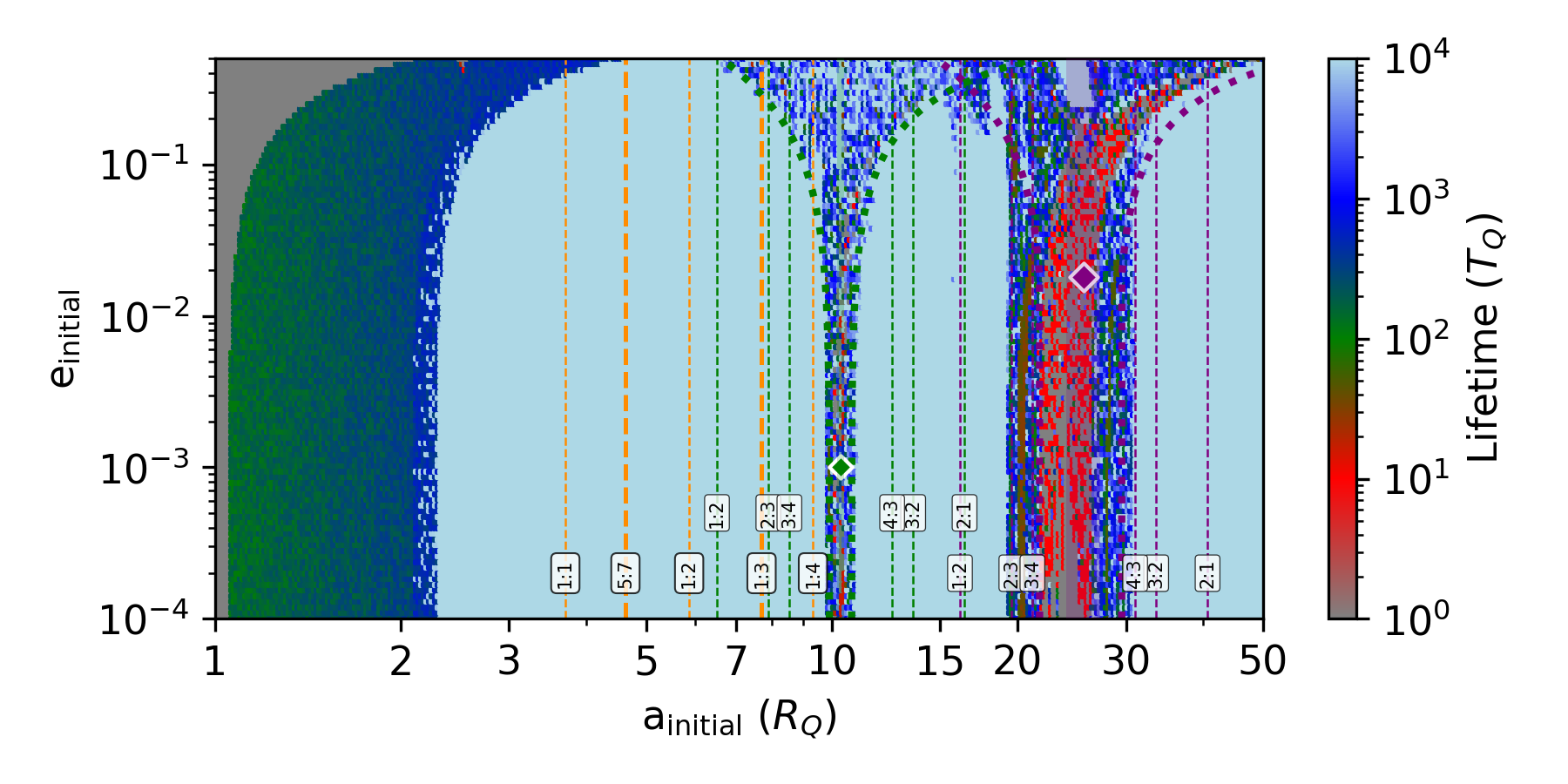}{0.49\textwidth}{(a)}
    \fig{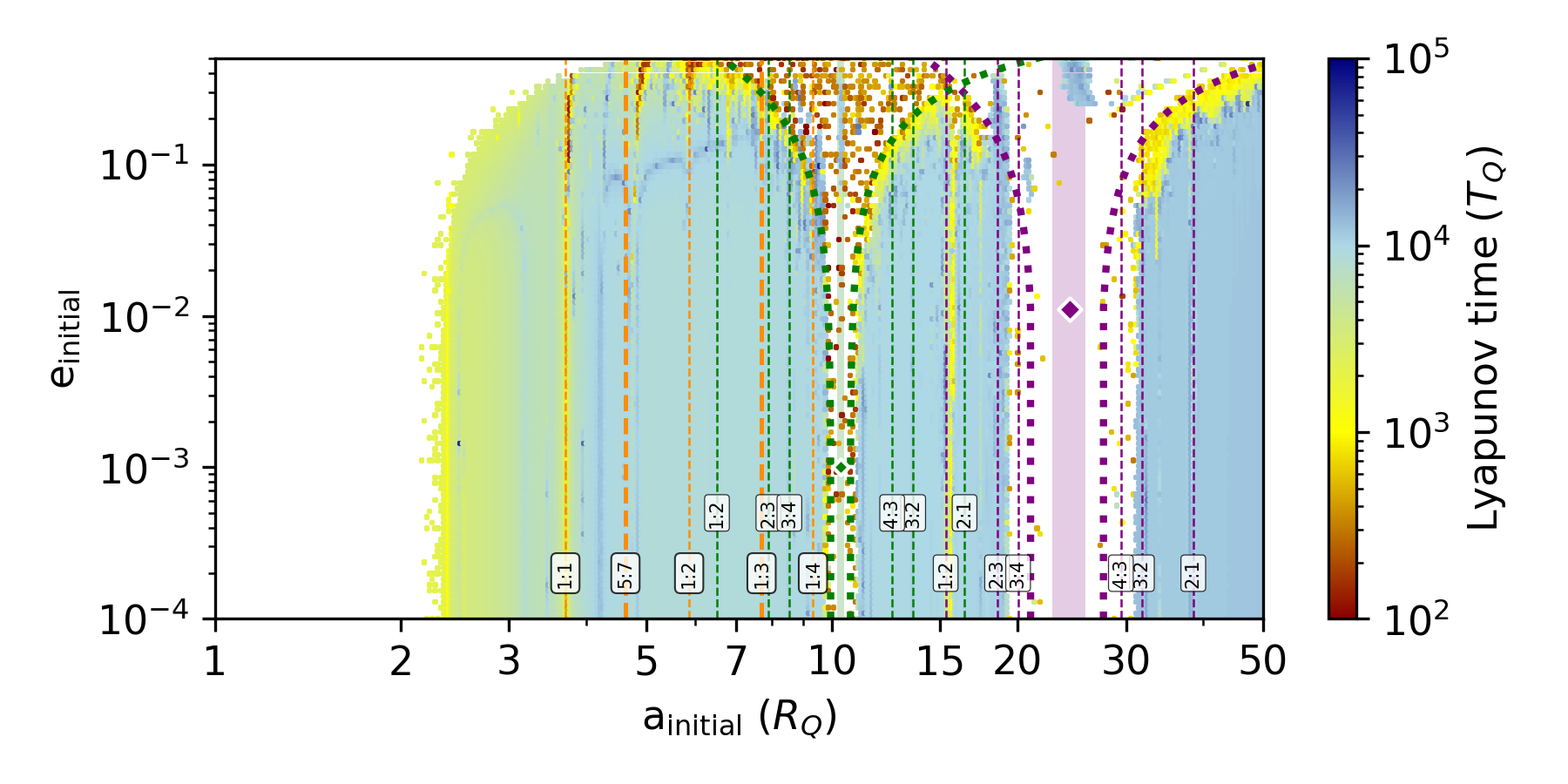}{0.49\textwidth}{(b)}}
  \caption{(a) Lifetime of all simulated particles and (b) Lyapunov time of the surviving particles, both as functions of the initial semi-major axis and eccentricity. Coloured vertical lines mark the locations of mean-motion resonances, with colours corresponding to those of the associated satellites (diamond markers). Orange vertical lines indicate spin–orbit resonances with Quaoar, with the wider ones highlighting the SORs associated with the rings.}
\label{fig:map}
\end{figure*}

In Figure~\ref{fig:4panels}, we present the evolution of the system by showing, at different times, the instantaneous semi-major axes and eccentricities of the particles (black dots) and satellites (diamond markers), whereas the particle stability as a function of their initial orbital elements is assessed in Figure~\ref{fig:map}. In Figure~\ref{fig:map}a, we show the particle lifetimes, while panel (b) presents the Lyapunov times of the particles that survive over the simulation timespan.

In both figures, the green and purple lines indicate the MMRs with the putative satellite and Weywot, respectively, while the orange lines mark the SOR locations. These resonance locations were computed assuming simple frequency commensurabilities and do not account for the effect of Quaoar’s shape, which may slightly shift the resonance positions, particularly those of the SORs \citep{Ribeiro2023,Madeira2025}.

The triaxial shape of an object is known to distort the trajectories of orbiting bodies into elliptical orbits in which the central object occupies the centre of the orbit \citep{Ribeiro2021}. In terms of osculating orbital elements, this effect translates into a \rv{induced} eccentricity in the orbiting material \citep{Winter2019,Ribeiro2023,Madeira2025}. In this context, the putative satellite acquires an eccentricity of \rv{$\sim 0.002$} due to the combined effects of Quaoar’s shape and the perturbation from Weywot. Particles with $a \lesssim 3.1~R_{\rm Q}$ are observed to reach minimum eccentricities of $\sim10^{-3}$ due to Quaoar's shape, with a large fraction achieving mean values of \rv{$\sim0.008-0.01$}. \rv{Such eccentricities} lead to close encounters with Quaoar and the putative satellite, resulting in chaotic diffusion of eccentricities and unstable motion \citep{Lages2017,Madeira2022a}.

Particles with initial semi-major axes up to \rv{$2.2~R_{\rm Q}$} collide with Quaoar on timescales of $\sim500~T_{\rm Q}$. Those with initial $a = 2.2$–$3.1~R_{\rm Q}$, in turn, are sufficiently distant from Quaoar to avoid collision over the simulation timespan; however, they exhibit shorter Lyapunov times than particles on more external orbits, which suggests that they may be removed on longer timescales or under the influence of perturbations not included in our model, like the presence of a small, undetected satellite in the region.
\begin{figure}
  \gridline{
    \fig{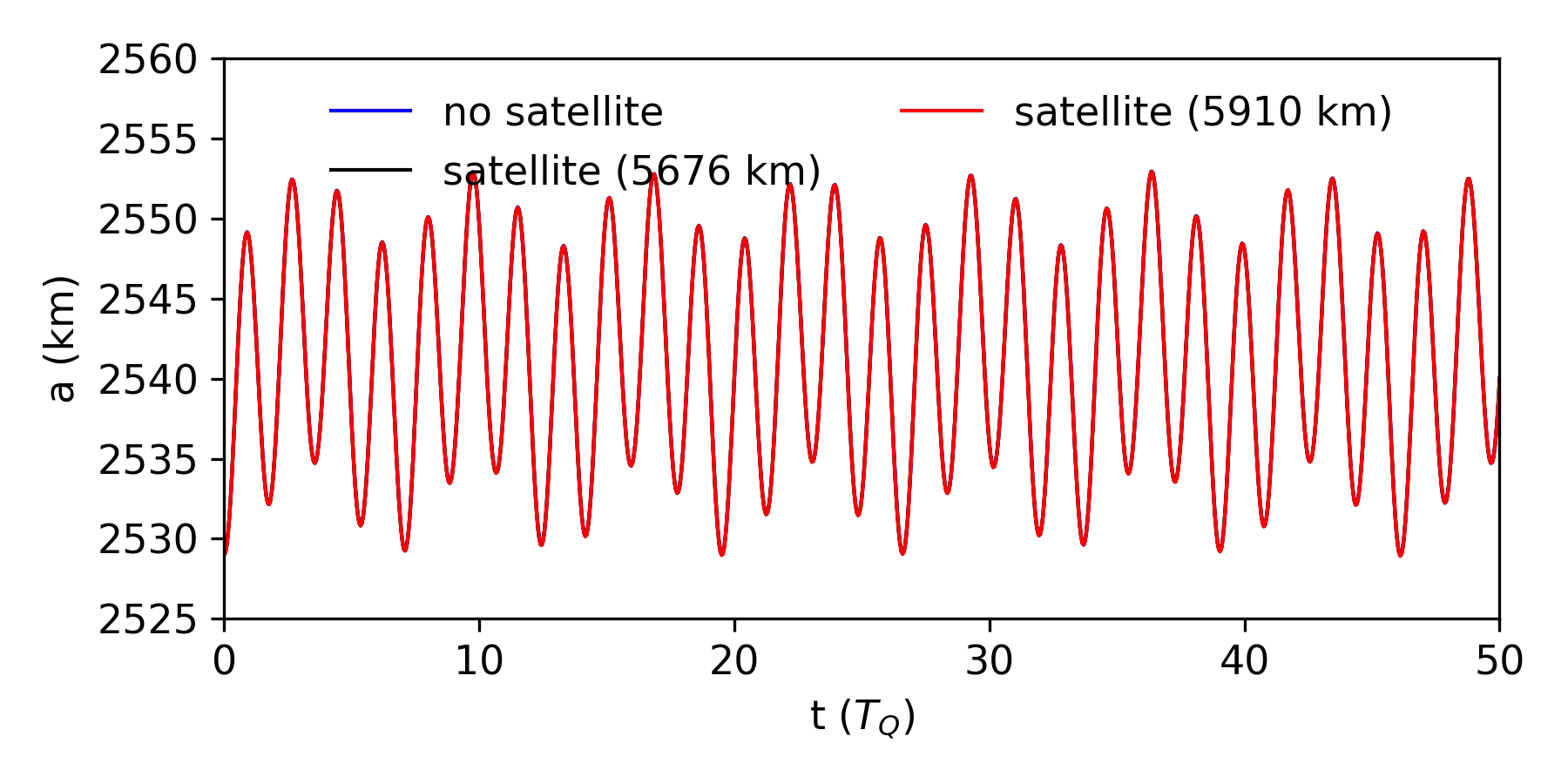}{\columnwidth}{(a)}}
\gridline{
    \fig{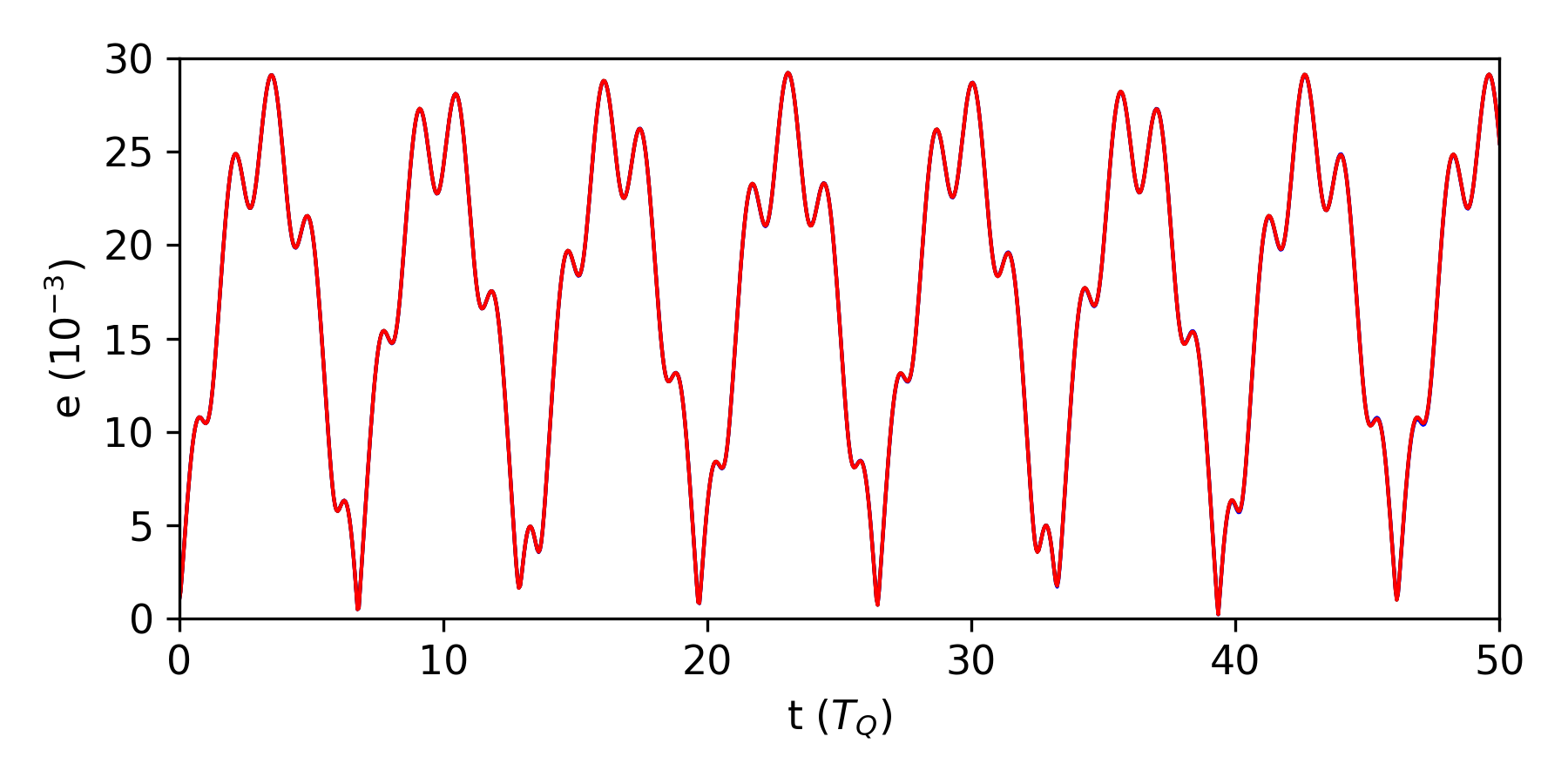}{\columnwidth}{(b)}
  }
\gridline{
    \fig{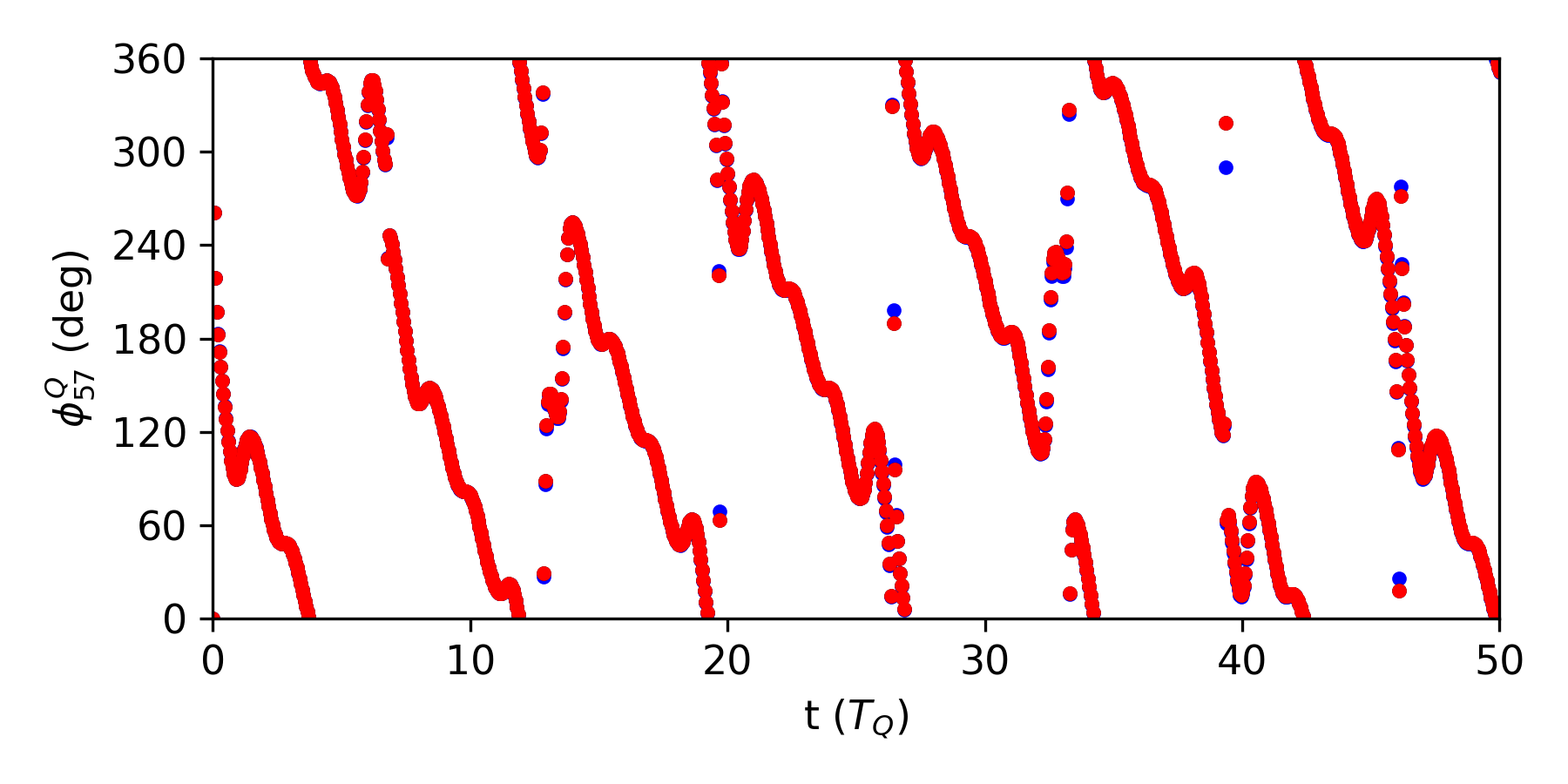}{\columnwidth}{(c)}
  }  
  \caption{\rv{Time evolution of the (a) semi-major axis, (b) eccentricity, and (c) 5:7 SOR resonant angle of a test particle initially placed at the radial centre of Q2R. The blue curves correspond to the configuration without the putative satellite, while the black and red curves correspond to configurations including a putative satellite at 5676 km and 5910 km, respectively. The resonant angle is defined as $\phi^{Q}_{57}=7\lambda-\frac{10\pi}{T_Q}t-2\varpi$, where $\lambda$ and $\varpi$ denote the particle mean longitude and longitude of pericentre, respectively.}}
  \label{fig:ringparticles_2}
\end{figure}

\begin{figure*}
  \gridline{
    \fig{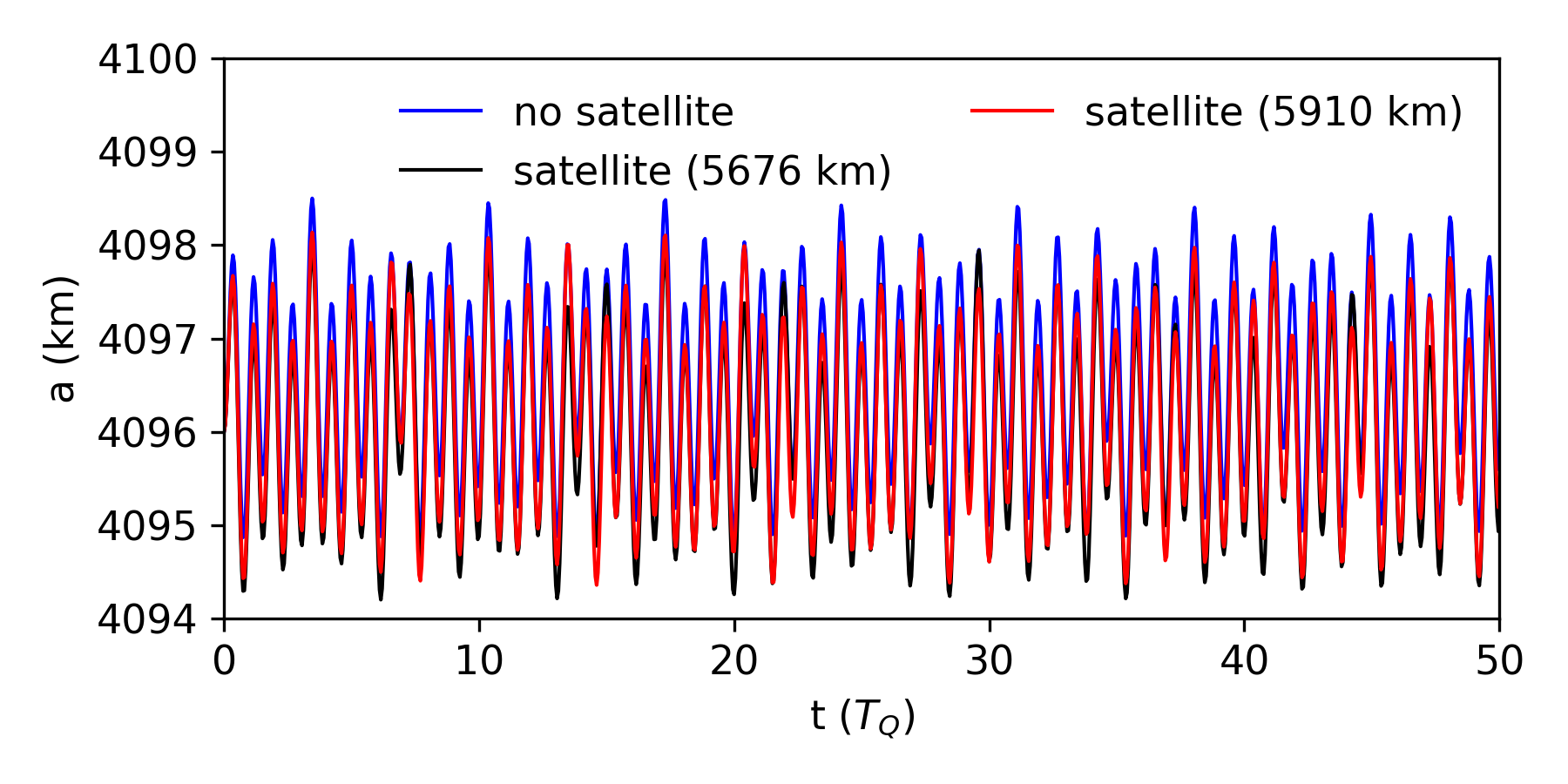}{\columnwidth}{(a)}
    \fig{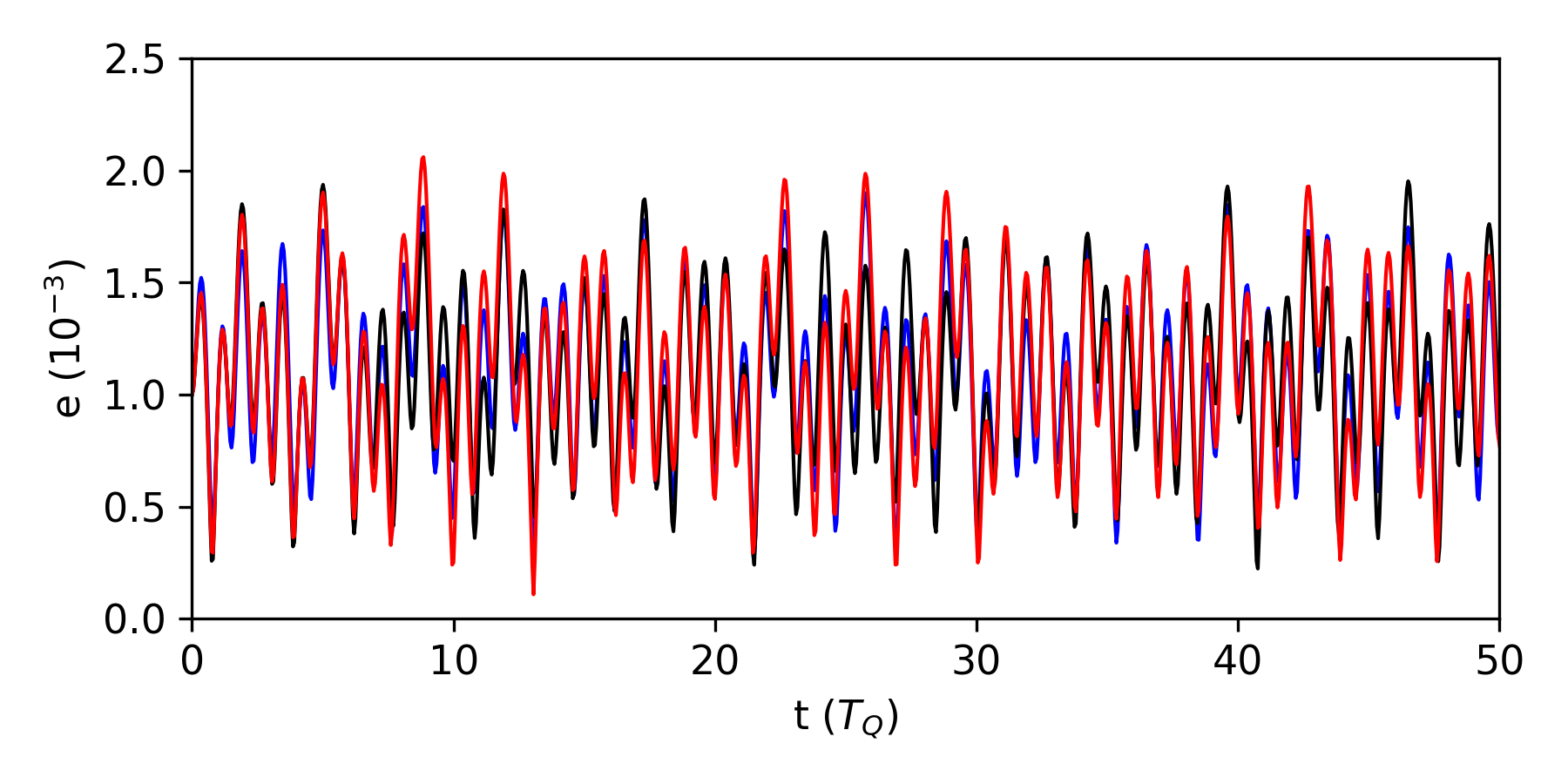}{\columnwidth}{(b)}
  }
\gridline{
    \fig{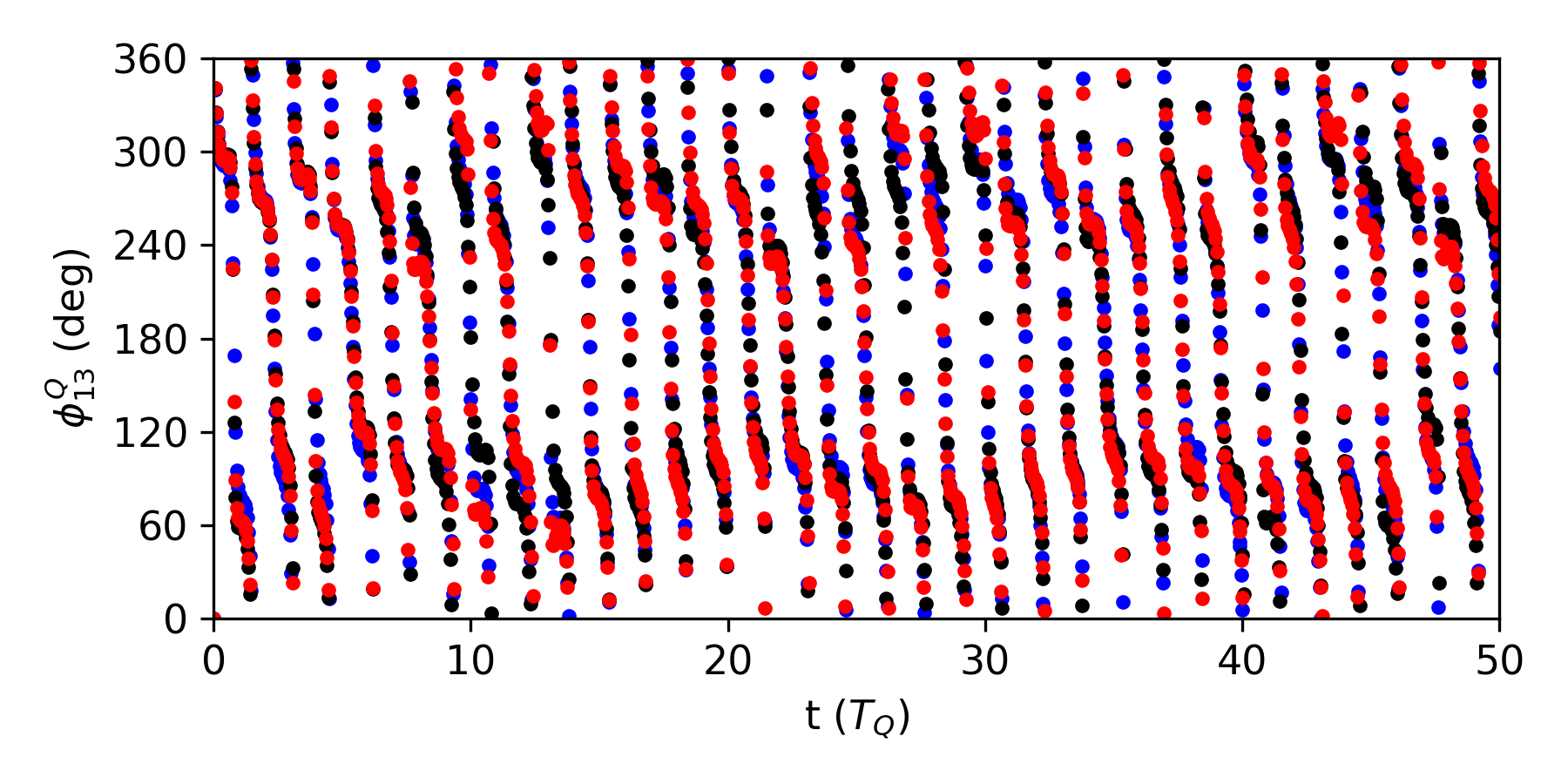}{\columnwidth}{(c)}
    \fig{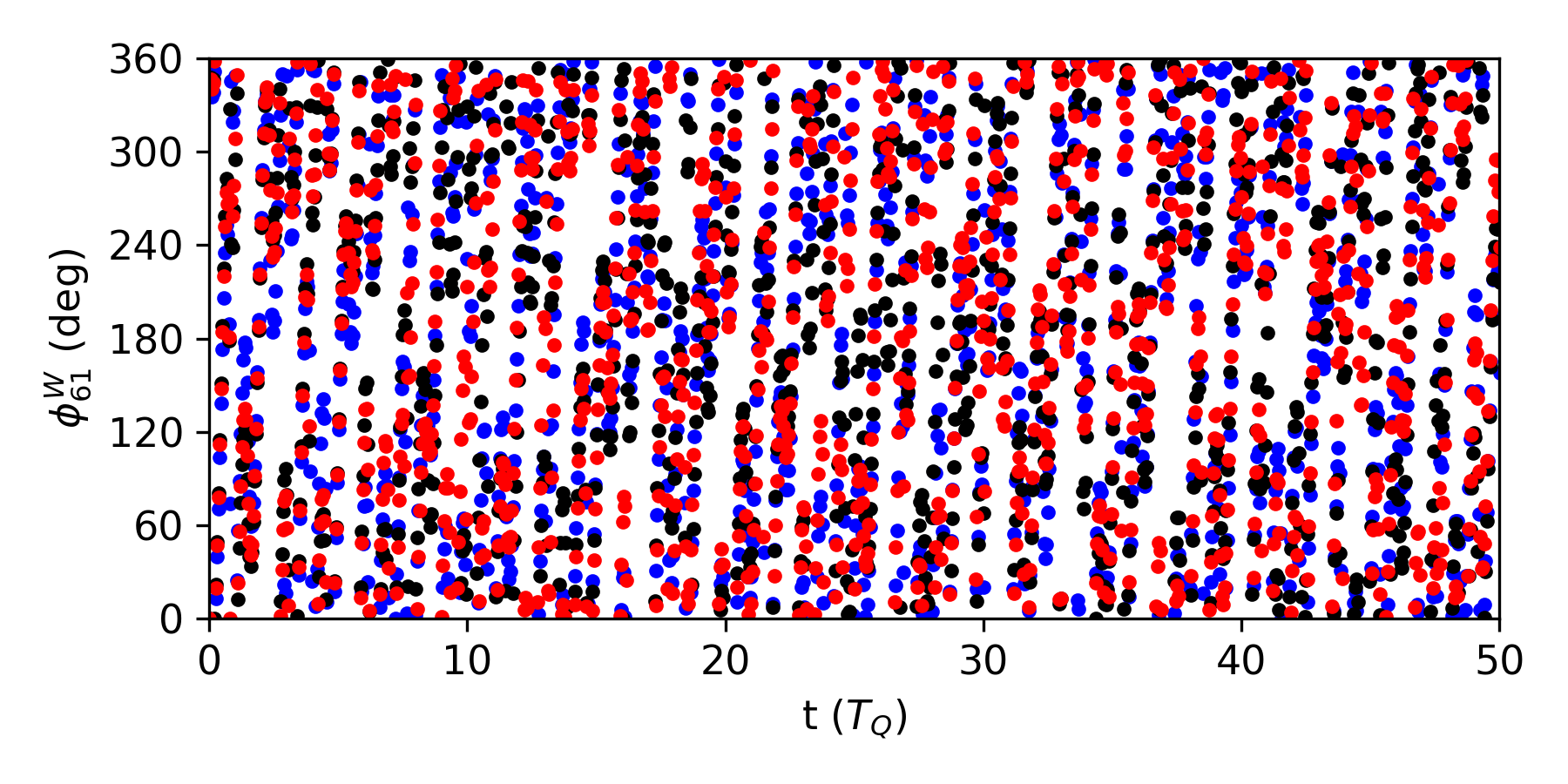}{\columnwidth}{(d)}
  }  
\gridline{
    \fig{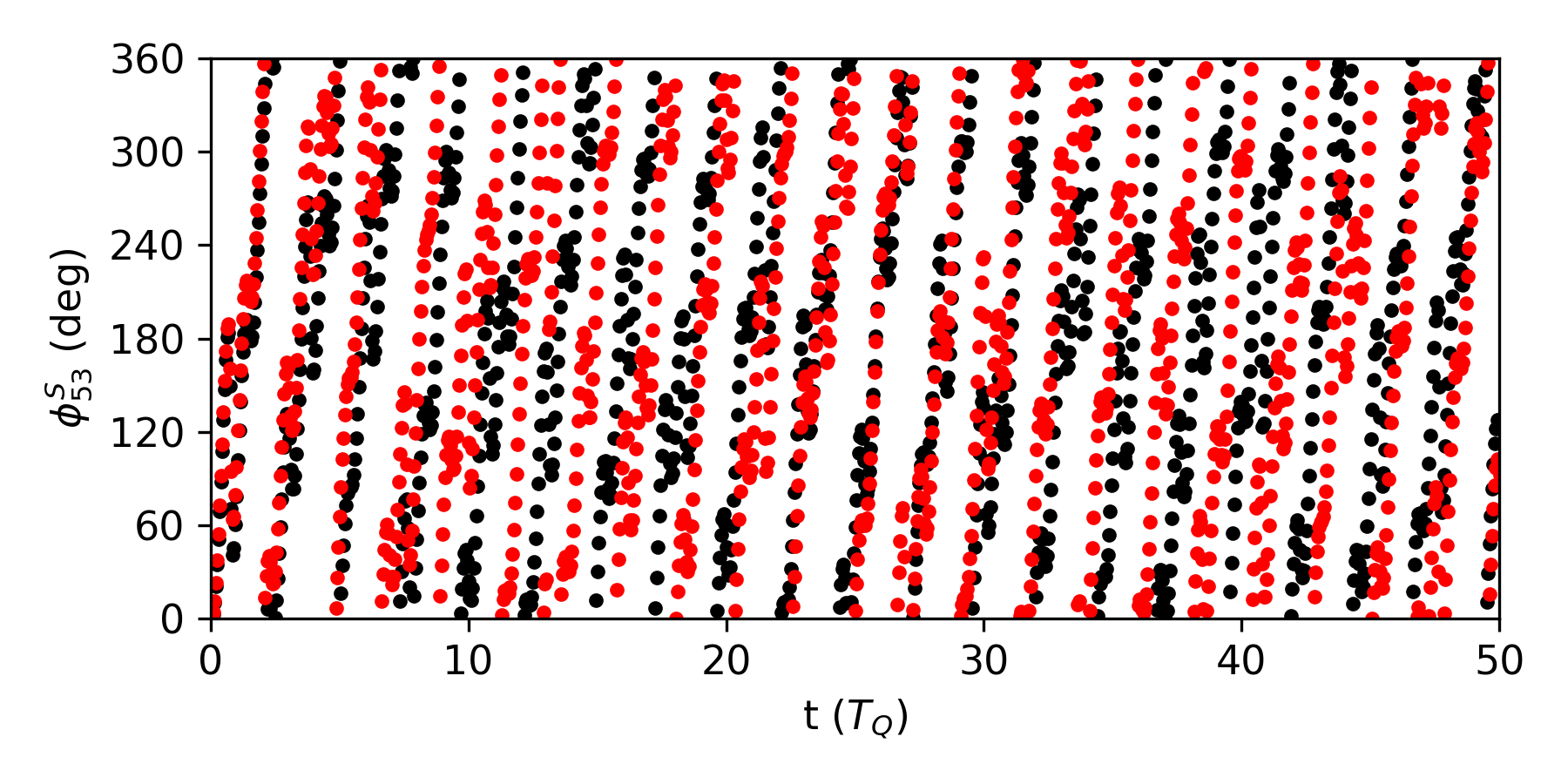}{\columnwidth}{(e)}
  }  
  \caption{\rv{Temporal evolution of the orbital elements and resonant angles of a test particle initially placed at the radial centre of Q1R. Blue curves correspond to the configuration with only Weywot in the system, black curves to the configuration including a putative satellite at 5676 km, and red curves to the configuration including a putative satellite at 5910 km. Panels (a) and (b) show the semi-major axis and eccentricity evolution, respectively. Panels (c), (d), and (e) show the resonant angles of the 1:3 spin–orbit resonance with Quaoar, the 6:1 mean-motion resonance with Weywot, and the 5:3 mean-motion resonance with the putative satellite, respectively. The resonant angles are defined as $\phi^{Q}_{13}=3\lambda-\frac{2\pi}{T_Q}t-2\varpi$, $\phi^{W}_{61}=6\lambda-\lambda_w-5\varpi$, and $\phi^{S}_{53}=5\lambda-3\lambda_s-2\varpi_s$, where $\lambda$, $\lambda_w$, and $\lambda_s$ denote the mean longitudes of the particle, Weywot, and the putative satellite, respectively, and $\varpi$ and $\varpi_s$ are the longitudes of pericentre of the particle and putative satellite, respectively.}}
  \label{fig:ringparticles_1}
\end{figure*}

In specific regions, particles are affected by Quaoar through SORs. This behaviour is observed near the 5:7, 1:2, and \rv{1:1} SORs, \rv{with the regions surrounding these resonances presenting a local decrease in particle concentration in Figure~\ref{fig:4panels}. We also see that particles in the vicinity of these resonances are less stable, exhibiting shorter Lyapunov times than those in nearby regions, particularly for particles with initial eccentricities of $\gtrsim0.1$.}

This result is expected for the 1:1 SOR and largely reflects our choice of initial conditions. A system with a rotating ellipsoidal central body possesses four equilibrium points that define the 1:1 SOR region \citep{Scheeres1994}. Two of these, aligned with Quaoar’s largest axis and located at \rv{$3.69~R_Q$}, are unstable saddle points, while the other two, aligned with the second-largest axis and located at \rv{$3.68~R_Q$}, are stable. Owing to our choice of initial longitude of pericentre, the particles are initially located near one of the saddle points and are therefore expected to be unstable. It remains to be investigated whether initial conditions with $\varpi \approx \pm 90^\circ$ would allow long-term surviving material in this region, which lies beyond the scope of the present work.

The result that \rv{high-eccentricity} material in the vicinity of the 5:7 SOR is less stable is particularly relevant, given that this resonance has been proposed as a confinement mechanism for Q2R \citep{Pereira2023}. Figure~\ref{fig:ringparticles_2} shows the evolution of the semi-major axis, eccentricity, \rv{and 5:7 SOR resonant angle ($\phi_{57}^Q$) of a representative particle with initial eccentricity of $10^{-4}$, placed at the centre of Q2R}. \rv{The results are presented for three different dynamical scenarios: one considering only Weywot as a satellite (in blue), one including the putative satellite at 5676~km (in black; nominal simulation), and one including the same satellite at 5910~km (red curves).}

\rv{As a first result, we find only negligible differences between the three scenarios. This occurs because the dynamical evolution of the Q2R region is dominated by Quaoar, while Weywot and the putative satellite exert only minor perturbations on the particles. Due to Quaoar, the particle reaches eccentricity of $\sim 0.03$, which is one order of magnitude greater than the benchmark value, indicating that Quaoar's ellipticity alone should be sufficient to induce velocity dispersions capable of preventing Q2R coagulation.} 

\rv{Assuming the reported radial location of the ring \citep[2529~km,][]{Proudfoot2025ring}, we obtain that the Q2R particle is not trapped in the 5:7 SOR, since the resonant angle does not librate. Nevertheless, its slow circulation and short-period oscillations around the resonance equilibrium points ($90^{\circ}$ and $270^{\circ}$) indicate that the particle is still affected by the resonance.}

In contrast to particles associated to inner SORs, those in the vicinity of the 1:3 SOR do not exhibit differences in lifetime or Lyapunov time when compared with nearby particles. \rv{In Figure~\ref{fig:ringparticles_1}, we also show the evolution of the semi-major axis, eccentricity, and relevant resonant angles of a particle initially placed at the reported radial location of Q1R \citep[4096~km,][]{Proudfoot2025ring}, for the same three dynamical scenarios. Panel~c shows the resonant angle of the 1:3 SOR ($\phi_{13}^Q$), panel~d the resonant angle of the 6:1 MMR with Weywot ($\phi_{61}^W$), and panel~e the resonant angle of the 5:3 MMR with the putative satellite ($\phi_{53}^S$). The 6:1 MMR with Weywot has been proposed as a possible confinement mechanism for Q1R if Weywot has a significant eccentricity \citep[$e\sim0.1$,][]{Rodriguez2023}, while the 5:3 MMR with the putative satellite may also lie near Q1R according to the findings of \citet{Proudfoot2025a}.}

\rv{For the Q1R particle, the putative satellite is observed to have a more prominent effect, affecting the oscillations of both the particle semi-major axis and eccentricity. Nevertheless, such variations are not responsible for inducing significant changes in the particle evolution. The particle reaches eccentricities of $\sim 0.001$, of the same order as the benchmark value, indicating that Quaoar's ellipticity, together with the contribution from Weywot, should still be sufficient to induce eccentricities in Q1R capable of preventing coagulation. Regarding the resonant angles, we observe that the particle at the reported Q1R location is not trapped in any of the resonances, although $\phi_{13}^Q$ presents slow circulation and short-period oscillations around the resonance equilibrium points, indicating an influence of the 1:3 SOR.}

\rv{As a general result, we observe that the effects of the putative satellite and Weywot become more relevant in comparison to Quaoar's influence for particles with initial semi-major axes greater than $\sim7~R_{\rm Q}$}, with particle eccentricities increasing as their orbits approach \rv{those of} the putative satellite and Weywot. The satellites are observed to clear, on timescales of ${\rm \sim 10^2–10^3~T_Q}$, most of the particles whose apocentres and pericentres cross the chaotic zone in their vicinity (dotted non-vertical lines in Figure~\ref{fig:4panels}). This region results from the overlap of first-order MMRs with the satellite and has a characteristic width given by
\begin{equation}
w_{\rm gap}=\iota\left(\frac{m_{\rm sat}}{M_Q}\right)^{2/7} a_{\rm sat},
\end{equation}
where $\iota = 2.6$ in the planar restricted three-body problem \citep{Wisdom1980}.

Here, we empirically obtain a lower value of \rv{$\iota = 1.8$}, a consequence of the particle eccentricities induced by Quaoar’s triaxial shape, which hinder the capture of particles in MMRs and thereby inhibit the extent of the chaotic region. We emphasise that particles surviving within the satellites’ chaotic zones exhibit short Lyapunov times of $\lesssim 10^3~T_Q$, confirming their chaotic nature, and are therefore expected to be eventually removed from the system.

We also verify the survival of stable particles in the coorbital regions of the satellites (vertical bands in Figures~\ref{fig:4panels} and \ref{fig:map}, $w_{\rm hs} \sim 0.5 (m_{\rm sat}/M_Q)^{1/3} a_{\rm sat}$), with associated Lyapunov times of $\gtrsim 10^4~T_Q$. Such particles are confined in horseshoe orbits and exhibit high eccentricities. For Weywot, we find that only particles with initial eccentricities of \rv{$\gtrsim 0.2$} survive in the coorbital region, whereas for the putative satellite, particles with initial eccentricities of $\gtrsim 0.05$ survive stably. It shows that, even if the detected feature is indeed a satellite, an associated arc or coorbital ring may still exist in the system.

In the region between $3.1$ and $15~R_{\rm Q}$, \rv{with the exception of the chaotic zone associated with} the putative satellite, we observe a general stability and survival of particles. This behaviour can \rv{also} be interpreted as an effect of Quaoar’s shape hindering the capture of particles in MMRs with the satellites, thereby avoiding large eccentricity increases associated with resonances and the consequent close encounters and loss of material. 

Although particles also generally survive in the region external to $15~R_{\rm Q}$, a more prominent effect of Weywot is observed there, given its higher mass compared to the putative satellite and its larger distance from Quaoar. \rv{A less stable region, characterized by lower Lyapunov times, is observed around} $15~R_{\rm Q}$, where the 1:2 MMR with Weywot overlaps the 2:1 MMR with the putative satellite. \rv{The same can be stated for the regions of overlap between the 2:3 and 3:4 MMRs and the region around the 4:3 MMR with Weywot, where particles exhibit shorter lifetimes.}

Given this, our results indicate that, if the feature is indeed a satellite, it is not expected to have a significant effect on material orbiting Quaoar outside its own chaotic region. So, other satellites, rings, or arcs may reside stably in its vicinity, as proposed by \citet{BragaRibas2026}. In Section~\ref{sec:undiscovered}, we explore the possibility of other satellites residing in the system. 

\rv{Another relevant result is that the putative satellite should have only a minor effect on the ring particles. Due solely to the ellipticity of Quaoar, we observe that Q1R and Q2R particles reach eccentricities of the same order as or greater than the benchmark value, which may explain the maintenance of these structures as rings. Nevertheless, at least at the reported radial locations of the rings, the particles are not trapped in any SOR, leaving the rings without confinement mechanism. Such confinement may instead be provided} by shepherd satellites near the rings \citep{GiuliattiWinter2023}, or by a stronger resonant mechanism than that caused by Quaoar's ellipticity, possibly associated with a mass anomaly \citep{Sicardy2026,Salo2026}.

\section{The feature as a coorbital arc} \label{sec:arc}

We now turn our attention to the scenario in which the feature corresponds to an arc. An important constraint of this interpretation is that the maintenance of an incomplete ring requires an azimuthal confinement mechanism, which, in principle, one could argue may be provided by SORs with Quaoar or by resonances with Weywot \citep{Goldreich1982,Scheeres1994}. However, the feature is observed to lie far from all first- and second-order SORs with Quaoar (Figure~\ref{fig:4panels}), while the closest resonance with Weywot (the 7:2 MMR, located about 100~km from the arc) is not expected to provide azimuthal confinement, as it is a fifth-order resonance \citep{BragaRibas2026}. In this context, the most plausible mechanism for the arc’s confinement is resonant forcing by an undiscovered satellite. 

We therefore explore the simplest configuration, in which the arc is trapped around one of the stable triangular equilibrium points of the Quaoar–satellite–particle system. In particular, we search for the satellite conditions capable of reproducing the dimensions of the putative arc, namely a radial extent of $23 \pm 2$~km and an angular width smaller than $28^{\circ}$ \citep{BragaRibas2026}. To this end, we performed a set of numerical simulations including an ellipsoidal Quaoar, Weywot, the confining satellite -- initially placed on a circular orbit at 5676.5~km -- and 100 test particles, while varying the mass of the satellite.

For each set of simulations, we determined the triangular equilibrium point analogous to the classical $L_4$ through dedicated test integrations, in which test particles were initially placed on coorbital trajectories with the satellite, at azimuthal separations of approximately $60^{\circ}$ from it. We then identified the angular location corresponding to the largest radial excursions of the particles \citep[see][]{Renner2004,Madeira2022}. For all considered satellite masses, we obtain displacements smaller than $\sim 0.3^{\circ}$ relative to the classical $L_4$ location. The 100 test particles used in the main simulations were then distributed uniformly within an angular sector defined by $-0.5 \leq r \leq 0.5$~km and $-1 \leq \theta \leq 1^{\circ}$ around the determined equilibrium point.

\begin{figure*}
  \gridline{
    \fig{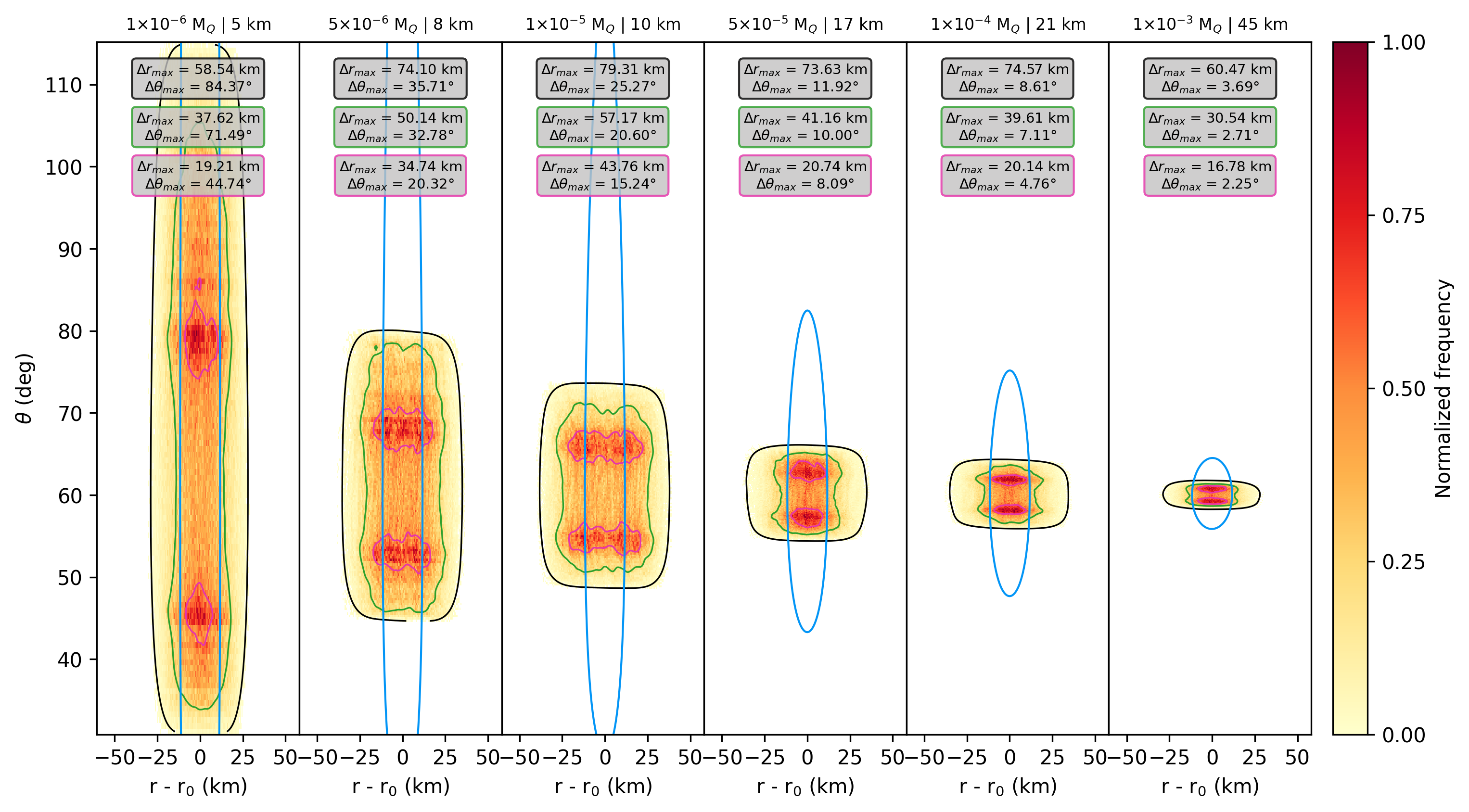}{1.0\textwidth}{}
    }
\caption{Frequency maps of test particles in the vicinity of one of the triangular equilibrium points of a satellite \rv{placed at 5676.5~km from Quaoar}, with the satellite mass indicated at the top of each panel. We consider masses ranging from $10^{-6}$ to $10^{-3}~M_Q$, corresponding to satellite radii of approximately 4 to 45~km. The radial origin of each map is defined by the median radial distance of the particles, $r_0$. The colour scale represents the normalized particle frequency. Black, green, and magenta contour lines delineate regions where the normalized frequency exceeds 0, 0.3, and 0.75, respectively. For each contour level, the grey boxes indicate the maximum extent of the particle distribution, quantified by $\Delta r_{\max}$ and $\Delta \theta_{\max}$. Blue curves correspond to zero-velocity curves in the restricted three-body problem that reproduce the radial width of the putative arc.}
\label{fig:arc}
\end{figure*}

Figure~\ref{fig:arc} shows frequency maps of test particles around the triangular equilibrium point after $60\,000~T_Q$, for satellite masses ranging from $10^{-6}$ to $10^{-3}~M_Q$. The first $10\,000~T_Q$ are discarded in order to minimize the influence of the initial conditions on the frequency maps. The maps are centred on the median radial distance of the particles, $r_0$, which does not coincide with the initial orbital radius of 5676.5~km. This offset arises from the eccentric orbit acquired by the satellite due to the combined effects of Quaoar’s shape and Weywot’s perturbation. For the smallest satellite masses, displacements of up to $\sim$50~km are obtained, as a consequence of eccentricities of order $\sim 10^{-2}$ attained by the satellite. For the case with $10^{-3}~M_Q$, for which the satellite is less susceptible to Weywot’s perturbation, the acquired eccentricity is only $\sim 10^{-4}$, resulting in radial displacements of just a few kilometres.

Black contours delimit regions where particles are present in at least one output. The regions enclosed by the green and magenta contours correspond to areas where particles are preferentially concentrated, with normalized frequencies exceeding 0.3 and 0.75, respectively. The gray boxes indicate the maximum radial and angular extents, quantified by $\Delta r_{\max}$ and $\Delta \theta_{\max}$, for each contour level. For reference, the blue curves show the zero-velocity curve in the circular restricted Quaoar–satellite–particle problem that reproduces the radial width of the arc. 

As a rule, we find that larger satellites provide stronger angular confinement, in agreement with predictions from the restricted three-body problem. Considering the regions enclosed by the black contours -- which would correspond to a sharp-edged arc, as suggested by the light-curve drops during the occultation -- we find that minimum masses of $10^{-5}~M_Q$ are required to produce an arc with an angular width consistent with the observations. This is one order of magnitude larger than the mass proposed by \cite{BragaRibas2026}.

However, no single satellite is able to reproduce the proposed radial confinement of the arc. As seen in Figure~\ref{fig:arc}, the black contours largely exceed the blue curves in the radial direction in all cases. Even when considering an arc with more diffuse edges, represented by the green contours -- which is not fully consistent with the observations -- the required radial width is not reproduced for the explored satellite masses.

Such inability of the satellites to successfully reproduce the radial extent of the feature is verified to be a direct consequence of Weywot’s perturbation, which induces large eccentricity variations in both the satellite and the particles. The satellite’s forced eccentricity then induces additional particle excitation, increasing the radial spreading of the arc.

\rv{Another direct consequence of Weywot's perturbation is revealed by tests performed with different satellite locations. In the case where the satellite is placed farther from Quaoar, at 5910~km (and therefore closer to Weywot), the radial confinement becomes even less efficient. Conversely, simulations with the satellite located closer to Quaoar show an improvement in the radial confinement of the particles.}

We confirm this interpretation by rerunning the simulations without Weywot. In its absence, the confinement becomes more consistent with that expected from the classical restricted three-body problem (blue curves), and arcs compatible with the observational constraints are obtained for satellite masses between $5\times 10^{-6}$ and $10^{-3}~M_Q$. By inspecting the resonant angles associated with the 7:2 MMR with Weywot -- both for the satellite and for the particles -- we find no cases of libration. We therefore rule out this resonance as the primary driver of the eccentricity excitation.

\rv{In the presence of Weywot, and with the satellite located at 5676.5~km}, the radial width of the arc is reproduced only for satellites with masses comparable to, or larger than, that of Weywot. In such cases, however, the satellite itself would have a cross section larger than that of the arc, making it unlikely that the arc would be detected through occultation while the satellite remains unseen. Moreover, a satellite of this size would lie at the threshold of detectability by direct imaging (Section~\ref{sec:constraint}). We therefore conclude that, if the feature corresponds to an arc, it cannot be explained by the confinement mechanism explored here.

\section{Undiscovered moons} \label{sec:undiscovered}
We now turn our attention to the possibility that undiscovered satellites exist in the system, assessing through numerical simulations where such objects may reside. Each numerical simulation includes a triaxial Quaoar, Weywot, the satellite, a moon, and a set of particles representing each ring. The simulations end when the moon collides with another major body, accretes all the \rv{particles of a ring}, or reaches a timespan of $10^4~T_Q$.

The satellite is placed at a semi-major axis of 5676~km, with all other orbital elements set to zero. For each ring, 10 particles are uniformly distributed in semi-major axis across the radial extent of the ring, with their initial eccentricities set to \rv{$10^{-3}$} -- a typical eccentricity attained by surviving particles in Figure~\ref{fig:4panels} -- and with all angular elements set to zero. The semi-major axis of the moon is varied uniformly and logarithmically from \rv{$2.0~R_Q$} -- given that particles interior to this distance are lost -- out to Weywot’s semi-major axis, sampling a total of 100 values. The physical radius of the moon is also varied uniformly and logarithmically from 10~m to 100~km, again sampling 100 values. The moon is assumed to have the same bulk density as Quaoar, an initial eccentricity of \rv{$10^{-3}$}, and all other angular orbital elements set to zero.

\begin{figure*}
  \gridline{
    \fig{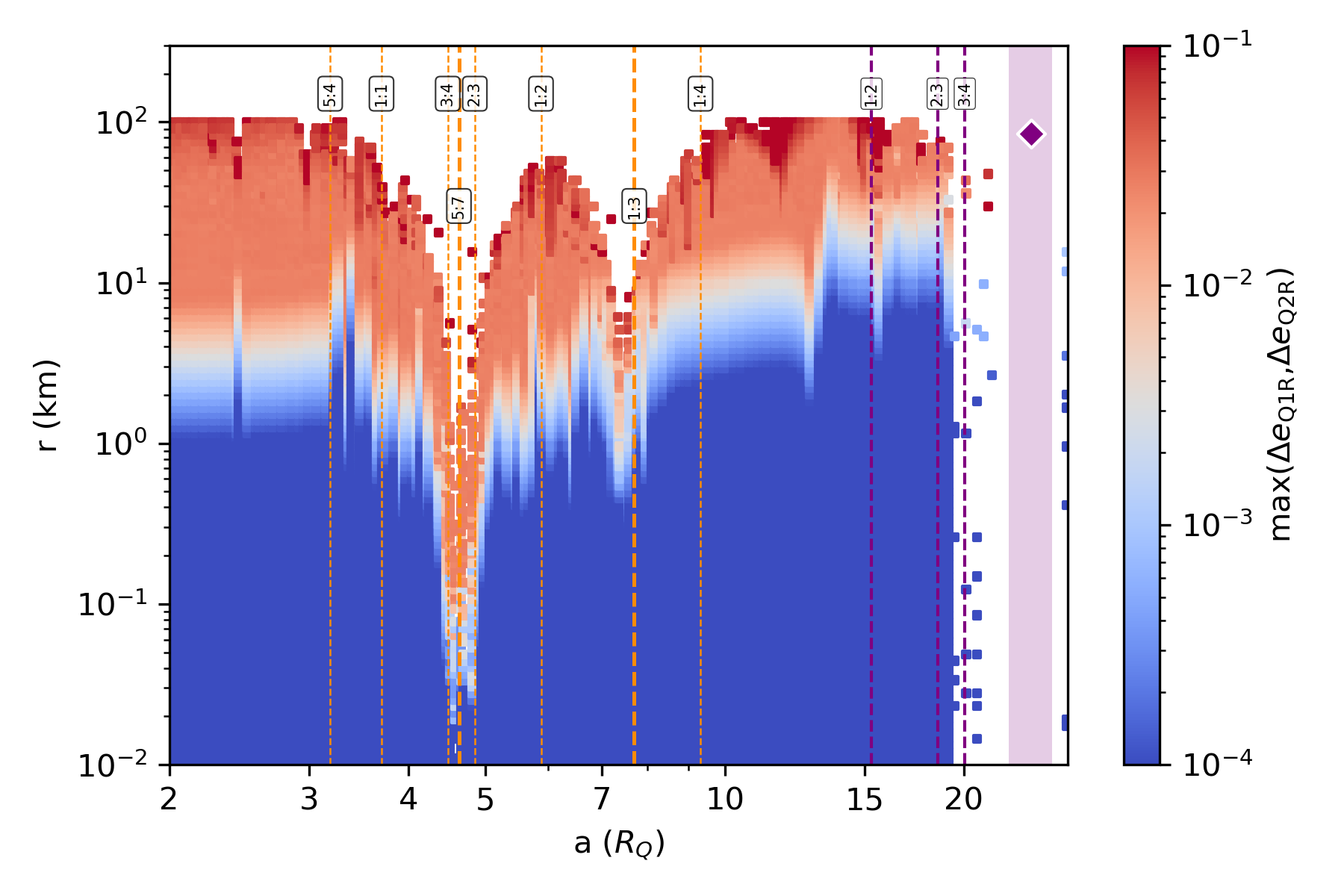}{0.49\textwidth}{(a) ${\rm 0}$}
    \fig{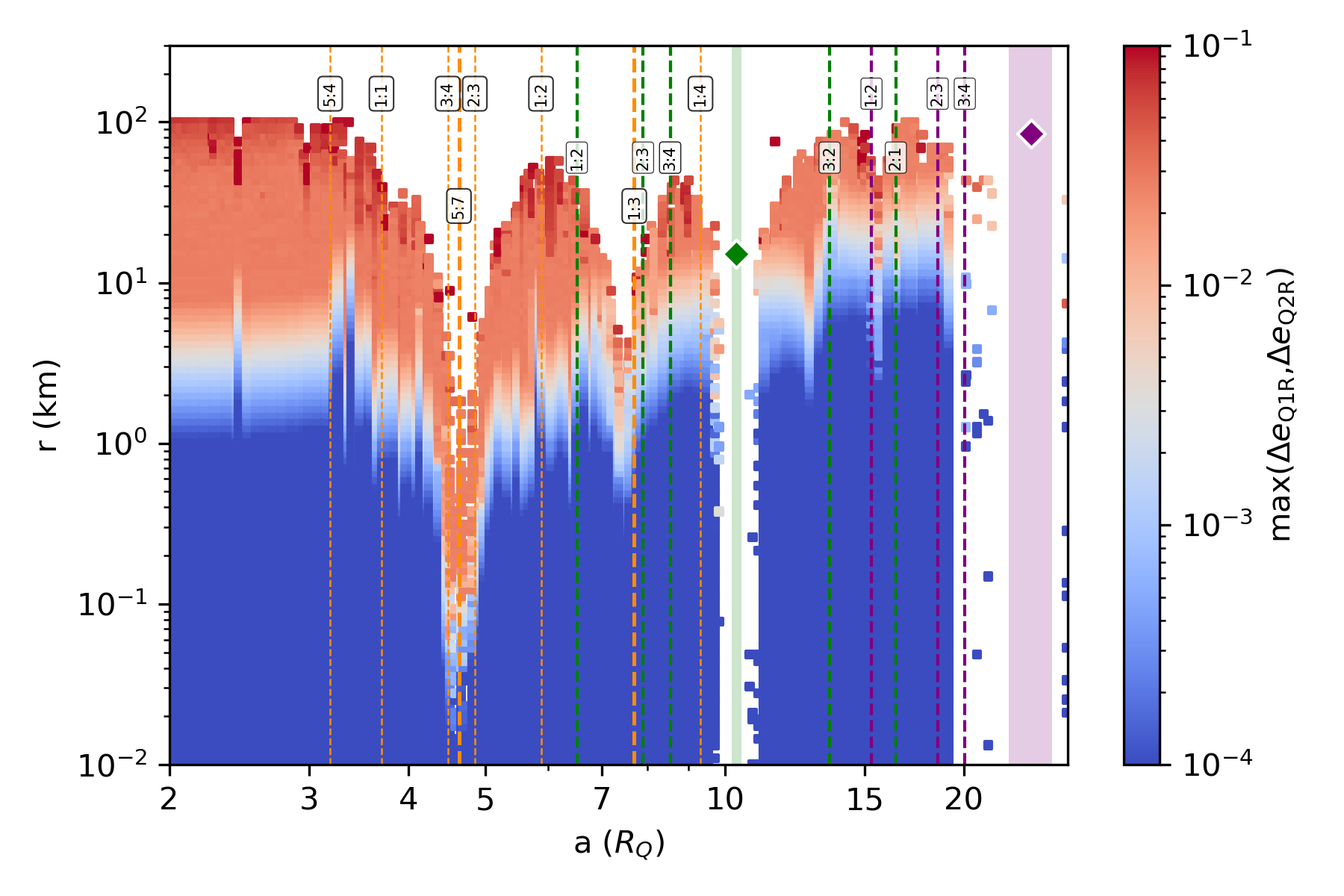}{0.49\textwidth}{(b) ${\rm 2\times10^{-5}~M_Q}$}
  }
  \gridline{
    \fig{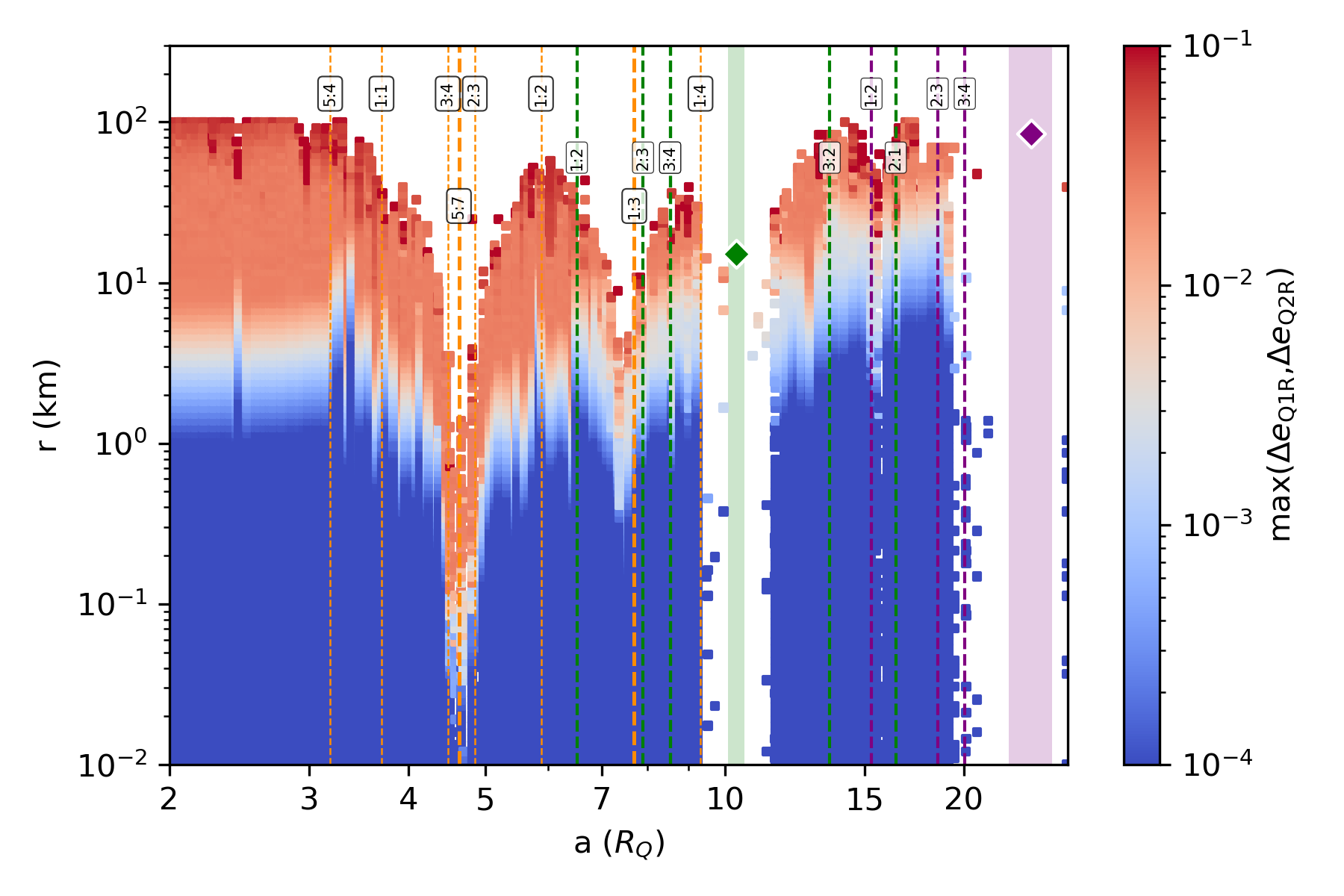}{0.49\textwidth}{(c) ${\rm 10^{-4}~M_Q}$}
    \fig{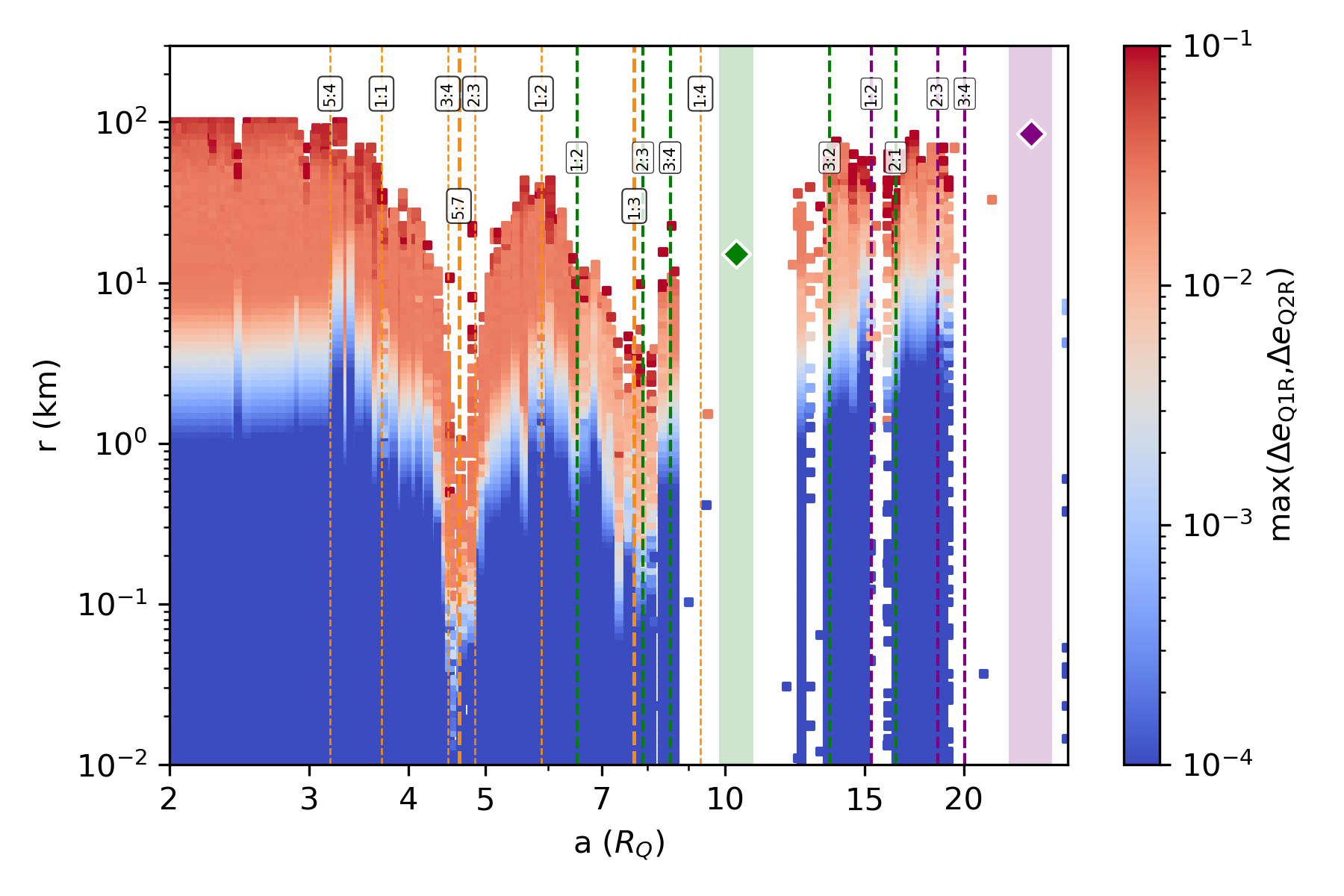}{0.49\textwidth}{(d) ${\rm 10^{-3}~M_Q}$}
  }
\caption{Initial semi-major axis and physical radius of a hypothetical additional moon that survives over $10^{4}~T_Q$ without accreting ring particles, with the colour scale indicating the maximum eccentricity induced by the moon on the ring particles. We consider the satellite associated with the feature with masses of (a) 0 (no satellite), (b) ${\rm 2\times10^{-5}~M_Q}$, (c) ${\rm 10^{-4}~M_Q}$, and (d) ${\rm 10^{-3}~M_Q}$. The hypothetical satellite (green diamond) and Weywot (purple diamond) are indicated by diamond markers, with their coorbital regions shown as vertical bands. SORs with Quaoar, MMRs with the satellite, and MMRs with Weywot are marked by orange, green, and purple vertical lines, respectively. The locations of the SORs associated with Q1R and Q2R are highlighted by wider dashed orange lines.}
\label{fig:2panels}
\end{figure*}

Figure~\ref{fig:2panels} shows the initial semi-major axis and physical radius of the moons in the simulations that reach the full simulation timespan, for different assumed satellite masses. For comparison, we explore the case without the satellite (panel~a), the case in which the feature directly corresponds to the satellite (panel~b), and two cases with a more massive satellite that could azimuthally confine the feature if it is an arc, although not radially (panels~c and d). In addition, we track the eccentricity variations induced by the additional moon on the ring particles; these are indicated by the colour scale, which represents the maximum eccentricity variation induced in particles from Q1R or Q2R.

For satellite masses $\lesssim 10^{-4}~M_Q$, the satellite exerts only a minor influence on the dynamical evolution of moons interior to Q1R. Its gravitational effect becomes significant only for masses $\gtrsim 10^{-3}~M_Q$, for which it induces large eccentricity variations in the moon, which in turn lead to enhanced eccentricity variations in the ring particles. 

\rv{Moons with sizes up to a hundred of kilometres located interior to the 1:1 SOR} are observed to survive over the entire simulation timespan. Within this stable region, a moon with a physical radius of $\sim3$~km is already sufficient to induce eccentricity variations in Q1R particles above the benchmark value.


In the immediate vicinity of the rings, large moons accrete all ring particles and therefore cannot reproduce the observed system. To reproduce the ring structures, moons with radii of up to $\sim5$~km may exist in this region, \rv{with objects as small as $\sim500$~m already capable of inducing eccentricity variations of the order of the benchmark value}. These results indicate that sub-kilometric shepherd moons may reside close to the rings, contributing to their confinement while enhancing impact velocities.

Beyond $8~R_Q$, the gravitational influence of the satellite and of Weywot becomes more prominent, with the regions surrounding these bodies being efficiently cleared of moons. Weywot removes \rv{almost all moons beyond $18~R_Q$, with only a few satellites surviving, possibly only temporarily, around the 3:4 MMR with Weywot.}

\rv{The satellite, in turn, clears moons from the region extending from about the 1:4 SOR ($\sim9~R_Q$) up to $\sim11~R_Q$ for satellite masses $\lesssim 10^{-4}~M_Q$, and from about the 3:4 MMR ($\sim8.5~R_Q$) to the location of the 3:2 MMR ($\sim13~R_Q$) for satellite masses $\gtrsim 10^{-3}~M_Q$. The existence of a cleared region at least $2~R_Q$ wide around the putative satellite location argues against the possibility of a satellite-belt around the object, as hypothesised by \citet{BragaRibas2026}, unless these satellites share the same orbit \citep{Madeira2022}.}

\rv{Between these regions cleared by the satellite and Weywot}, moons with sizes up to several tens of kilometres can remain stable. \rv{A local exception occurs for satellite masses of $\sim10^{-3}~M_Q$ in the overlap region between the 2:1 MMR with the satellite and the 1:2 MMR with Weywot}, where moons experience eccentricity growth driven by resonant interaction, eventually crossing the satellites’ orbits and being removed from the system.

\section{Discussion} \label{sec:conclusion}

In this work, we dynamically explored the two possible interpretations of the feature recently detected around Quaoar, as reported in \cite{Nolthenius2025} and in \cite{BragaRibas2026}: namely, a dense arc or a satellite. The arc scenario is statistically more probable, as such a structure would have a larger cross section, increasing the likelihood of producing a detectable signal in an occultation light curve. 

Nevertheless, the arc interpretation is dynamically more intricate, as it requires an azimuthal confinement mechanism, which is not expected to arise from the gravitational influence of either Quaoar or Weywot. Therefore, this scenario would necessarily imply the existence of at least one additional, yet undiscovered, satellite in the system.

Here, we explored the simplest known azimuthal confinement mechanism: a satellite trapping an arc around one of its triangular equilibrium points \citep{Lissauer1985}. Using frequency maps derived from numerical simulations, we modelled arcs confined by hypothetical satellites and compared the resulting radial and angular extents with those of the reported feature.

Although the satellites are capable of providing azimuthal confinement, we find that none with masses up to $10^{-3}~M_Q$ can reproduce the observed radial extent of the feature. This limitation arises from the gravitational influence of Weywot, which excites the eccentricities of both the satellite and the particles, largely increasing the radial excursions of the arc material. This behaviour is driven by the secular perturbation from Weywot rather than by the nearby 7:2 MMR, as confirmed by our analysis of the resonant angles associated with this resonance for both the satellite and the particles, which show no libration.

The frequency maps reveal significant temporal variations in particle density. The failure to reproduce the radial extent of the arc persists both for a sharp-edged configuration -- which better matches the observed light curve -- and for a more diffuse structure with evanescent edges. Only satellites large enough to be detectable by direct imaging are capable of reproducing the required radial confinement; such cases are therefore ruled out.

It is nevertheless interesting to note that satellite confinement leads to the formation of two high-density islands (enclosed by the magenta contours in Figure~\ref{fig:arc}) in the arc, implying that measurable variations in light-curve attenuation are expected due to particle accumulation within the structure. If embedded within a broader ring -- such as the C1R ring of Chariklo -- a satellite could generate a similar density pattern, potentially providing an explanation for the longitudinal opacity variations observed in Chariklo's system \citep{Morgado2021}. 

Given the above, we conclude that confinement of the arc at a triangular point of a satellite is not a realistic mechanism for explaining this new feature around Quaoar. We realise that our numerical simulations include only test particles and therefore do not capture the full complexity of a dense arc, as they neglect both self-gravity and inter-particle collisions. Nevertheless, given the very high optical depth ($\tau > 6$) inferred for the feature \citep{Nolthenius2025}, a high collision rate within the arc would be expected. Such conditions would likely lead to strong radial viscous spreading \citep[see][]{Madeira2025a,Madeira2025b}, making it difficult to maintain a sharp-edged structure. Alternatively, sustained collisional interactions beyond Quaoar’s Roche limit could favour accretion and satellite formation. Both scenarios argue against the long-term stability of the arc hypothesis.

We emphasise that, although the mechanism explored here is not successful in confining an arc with the observed extents, other confinement processes may operate in the system. Examples include confinement through a corotation eccentric resonance with an undetected satellite, or shepherding by a constellation of coorbital satellites \citep{Renner2004,Hedman2009,Madeira2020,Madeira2022}.

Nevertheless, regardless of the confinement mechanism invoked, the arc scenario still faces the fundamental difficulty of strong radial spreading. This is particularly problematic given that the occultation light curve indicates a sharp-edged structure. Alternatively, sustained collisional interactions may favour coagulation into a satellite rather than the long-term maintenance of a dense arc.

The most dynamically plausible interpretation is that the feature corresponds to a satellite, \rv{although \citet{BragaRibas2026,Proudfoot2025a} estimate the probability of detecting such an object through a single occultation event to be lower than $1\%$.}. This very low probability leads us to speculate that the Quaoar system may be significantly more complex than currently known, potentially hosting additional rings, arcs, and moons. Indeed, based on statistical arguments, \cite{BragaRibas2026} suggest that the number of kilometre-sized moons in the system could be as high as 36.

Under this scenario, we find that the putative satellite reaches an eccentricity of about \rv{0.002} due to the triaxial shape of Quaoar and the gravitational influence of Weywot. The satellite is found to have no significant dynamical effect on the Q1R and Q2R rings, inducing only minor variations in the eccentricities of ring particles \rv{for both radial locations proposed in the literature \citep{Proudfoot2025a,BragaRibas2026}}. Thus, it does not appear to play any relevant role in maintaining the rings as particulate structures, either by preventing coagulation or by providing confinement. Instead, its primary dynamical effect in the Quaoar system is to clear material in its immediate vicinity, creating a gap of approximately $\sim0.7~R_Q$ around its orbit.

Regarding the kilometre-sized yet-undetected moons that may reside in the system, our results indicate that:
\begin{itemize}
    \item Moons may reside between \rv{$2~R_Q$} and the location of the 1:1 SOR ($\sim3.7~R_Q$), with physical radii of up to \rv{$100$~km}. Even kilometre-sized moons in this region would be capable of inducing eccentricity variations of order \rv{$10^{-3}$} in Q2R particles.
    \item In the immediate vicinity of the rings, only moons with radii up to $\sim5$~km can stably reside. Bodies with radii larger than \rv{a few hundreds of metres would} already induce perturbations sufficient to inhibit coagulation within the rings.
    \item \rv{Moons with radii up to a few tens of kilometres can reside in the regions between $5$–$6~R_Q$ and $8$–$9~R_Q$, while the region between $9$–$11~R_Q$ should remain largely devoid of moons due to the effect of the putative satellite.}
    \item Between $11~R_Q$ and $20~R_Q$, large moons with radii of several tens of kilometres can stably reside, \rv{except in the vicinity of the overlap region between the 2:1 MMR with the satellite and the 1:2 MMR with Weywot, where material is dynamically less stable.}
\end{itemize}

We emphasise that our dynamical results depend on the adopted shape model of Quaoar, which is not yet fully constrained. The results obtained for the inner regions of the system should therefore be interpreted with caution, particularly those concerning Q1R and Q2R, which are strongly influenced by the body’s triaxiality. In contrast, at the radial location of the detected feature and beyond, Quaoar’s triaxiality has a minor dynamical effect, and our conclusions are therefore expected to be more robust against moderate variations in the primary’s shape.

\rv{For the currently adopted shape parameters \citep{margoti2024,Proudfoot2025_5}, inspection of the respective resonant angles shows that neither the Q1R nor the Q2R particles are trapped in the 1:3 and 5:7 SORs, respectively, although the resonant angles exhibit slow circulation and short-period oscillations around the resonance equilibrium points, indicating at least an influence of these resonances. Nevertheless, as highlighted, confinement through such resonances strongly depends on the adopted shape model and, given the uncertainties in the ring locations, our results do not imply that the rings are not truly in resonance. The same can be said for the 6:1 MMR between Q1R and Weywot, whose resonant angle is not observed to librate.}

\rv{One interesting result, however, is that even though the ring particles are not trapped in SORs in our simulations, which are known to excite particle orbits \citep{Ribeiro2023}, the non-resonant eccentricity increase induced by Quaoar's ellipticity alone is sufficient to generate velocity dispersions capable of maintaining a centimetre-sized icy ring against coagulation.}

These results, however, do not include collisional dynamics within the rings, which are known to play a fundamental role in confinement driven by SORs \citep{Salo2026}. Dedicated N-body simulations including collisions and self-gravity are therefore required to draw stronger conclusions regarding the origin and long-term maintenance of the Quaoar rings. 

Moreover, a single-chord occultation event cannot unambiguously distinguish between a small satellite, a very dense arc, or even a transient clump. This observational degeneracy can only be broken through multi-chord occultations or multiple detections in several events. Future stellar occultations, as well as long-term monitoring with facilities such as JWST or even the next-generation extremely large telescopes, will be crucial to unveiling the true architecture of the Quaoar system.

\begin{acknowledgments}
\rv{We thank both reviewers for their valuable comments and suggestions, which helped improve the quality of this manuscript.}
GM acknowledges the infrastructure provided by the Multiuser Data Processing Center of the National Observatory (CPDON). 
LE acknowledges funding from the Fundação de Amparo à Pesquisa do Estado de São Paulo (FAPESP) under grant 2021/00628-6.
BEM thanks CAPES grant 23079.212658/2024-30, CNPq/Universal grant 408543/2025-6 and FAPERJ grant E-26/204.205/2025.
SMGW thanks FAPESP (Proc.~2022/11783-5), CNPq (Proc.~309057/2025-6) and Capes (Financial Code~001). 
OCW thanks FAPESP (Proc.~2022/11783-5) and  CNPq (Proc.~316991/2023-6).
The numerical simulations were performed on the CDJPAS platform at National Observatory and on the Center for Scientific Computing (NCC/GridUNESP) of the São Paulo State University (UNESP).
During the preparation of this work, the authors used ChatGPT-5 in order to improve readability. After using this tool, the authors reviewed and edited the content as needed and take full responsibility for the content of the publication.
\end{acknowledgments}

\begin{contribution}
GM led the project, performed the dynamical simulations, and wrote the manuscript.
LE and PVSS performed the arc scenario simulations, analysed the results, and contributed specific parts of the manuscript.
BEM provided the observational constraints, analysed the results, and contributed specific parts of the manuscript.
SGW and OCW developed the manuscript concept, revised, and improved the manuscript.
BSC performed part of the dynamical simulations.
\end{contribution}

\bibliography{main}{}
\bibliographystyle{aasjournalv7}



\end{document}